\documentclass[12pt]{reportj}
\usepackage{deluxetablej}
\usepackage{hyperref} 
\makeatletter
\def\@to{to}
\makeatother
\usepackage{times}
\usepackage{graphicx}
\usepackage{tocloft}
\usepackage{fancyheadings}
\emergencystretch=\maxdimen
\usepackage{datetime}
\newdateformat{ddmonthyyyy}{\THEDAY\ \monthname[\THEMONTH]\ \THEYEAR}

\renewcommand\thesection{\arabic{section}}
\renewcommand\thesubsection{\thesection.\arabic{subsection}}

\def\ssection#1{\setcounter{subsection}{0} \refstepcounter{section} \section*{\hbox to \hsize{\large\bf \arabic{section}. #1\hfill }}\label{sec} \addcontentsline{toc}{section}{\arabic{section}. #1}}
\def\ssubsection#1{\setcounter{subsubsection}{0} \refstepcounter{subsection}\subsection*{\hbox to \hsize{\normalsize\bfseries\itshape \arabic{section}.\arabic{subsection} #1\hfill}}\label{subsec} \addcontentsline{toc}{subsection}{\arabic{section}.\arabic{subsection} #1}}
\def\ssubsubsection#1{\refstepcounter{subsubsection}\subsection*{\hbox to \hsize{\normalsize\it \arabic{section}.\arabic{subsection}.\arabic{subsubsection} #1\hfill}}\label{subsubsec} \addcontentsline{toc}{subsubsection}{\arabic{section}.\arabic{subsection}.\arabic{subsubsection} #1}}

\def\ssectionstar#1{\section*{\hbox to \hsize{\large\bf #1\hfill}} \addcontentsline{toc}{section}{#1}}
\def\ssubsectionstar#1{\subsection*{\hbox to \hsize{\normalsize\bfseries\itshape #1\hfill}} \addcontentsline{toc}{subsection}{#1}}
\def\ssubsubsectionstar#1{\subsection*{\hbox to \hsize{\normalsize\it  #1\hfill}} \addcontentsline{toc}{subsection}{#1}}

\renewcommand{\cftaftertoctitle}{%
\mbox{}\hfill{\normalfont Page}}
\begin{document}

~\\

\vspace{-2.4cm}
\noindent\includegraphics*[width=0.295\linewidth]{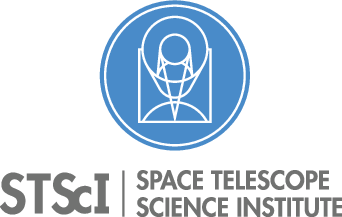}

\vspace{-0.4cm}

\begin{flushright}
    {\bf Instrument Science Report STIS 2025-05(v1)}
    
    \vspace{1.1cm}
    
    {\bf\Huge Status of the STIS Auto-wavecal Exposures}
    
    \rule{0.25\linewidth}{0.5pt}
    
    \vspace{0.5cm}
    
    D. Welty$^1$ and S. Lockwood$^1$
    \linebreak
    \newline
    \footnotesize{$^1$ Space Telescope Science Institute, Baltimore, MD}
    
    \vspace{0.5cm}
    
     \ddmonthyyyy{18 September 2025} 
\end{flushright}

\vspace{0.1cm}

\noindent\rule{\linewidth}{1.0pt}
\noindent{\bf A{\footnotesize BSTRACT}}

{\it \noindent 
We discuss the behavior of the default ``wavecal'' spectra obtained together with most STIS spectroscopic exposures, which are needed for proper wavelength calibration of the science data.
Because the Pt/Cr-Ne lamps used for the wavecals have been fading (especially at the shortest wavelengths), some changes in the default lamp and/or exposure time have been implemented in recent years to maintain accurate calibrations.
To assess whether additional changes might be appropriate, we examine the distribution of the {\small SHIFTA1, SHIFTA2} values derived from the wavecals (the x and y offsets of the spectral image on the detector), we re-visit the wavelength-dependent fading of the lamps, and we perform simulations to estimate the exposure times that would be needed to obtain accurate {\small SHIFTA} values.
While the current wavecals do appear to yield reasonable {\small SHIFTA}, increases in the default exposure times for some of the shortest-wavelength settings would help to ensure reliable wavelength zero points as the lamps continue to fade.}

\vspace{-0.1cm}
\noindent\rule{\linewidth}{1.0pt}

\renewcommand{\cftaftertoctitle}{\thispagestyle{fancy}}
\tableofcontents


\lhead{}
\rhead{}
\cfoot{\rm {\hspace{-1.9cm} Instrument Science Report STIS 2025-05(v1) Page \thepage}}



\vspace{-0.3cm}
\ssection{Introduction}\label{sec:Introduction}

Wavelength calibration of STIS spectroscopic exposures is accomplished via standard dispersion relations that map the (x,y) coordinates of the detector to (order,wavelength).
For each grating / central wavelength setting, those standard dispersion relations were established by fitting a simple dispersion formula (with up to cubic terms in $m$ and $\lambda$) to Pt/Cr-Ne emission-line spectra from deep pre-launch lamp exposures.
The various wavelength settings are implemented by using tabulated values for the slit wheel and the three Mode Selection Mechanism (MSM) cylinders.
Because the results of those standard MSM settings are not precisely reproducible, however, contemporaneous information is required to establish the exact location of the spectral image on the detector for any particular observation (\href{https://www.stsci.edu/files/live/sites/www/files/home/hst/instrumentation/stis/documentation/instrument-science-reports/_documents/199701.pdf}{Baum 1997}).
Almost all STIS spectroscopic observations are therefore accompanied by short ``wavecal'' exposures, taken at the same setting (and in many cases through the same aperture; Table~\ref{tab:aper}) with one of the three onboard Pt/Cr-Ne lamps.
The wavecals for most first-order MAMA and CCD settings use the HITM1 lamp; the wavecals for most FUV- and NUV-MAMA echelle settings use the LINE lamp.\footnotemark
\footnotetext{Initially, G140M science exposures taken through the wider 52x0.5 and 52x2 apertures were accompanied by wavecals obtained with the LINE lamp through the 52x0.1 aperture.  The lamp appears to have been switched to HITM1 for those cases in 2019.}
Those contemporaneous wavecals yield (x,y) zero point offsets (approximately along and perpendicular to the dispersion direction, respectively) for the associated science observations via cross-correlations with appropriate template images and/or spectra.
For echelle spectra, a 2D cross-correlation between the wavecal image and a corresponding template spectral image is performed; for first-order spectra, separate 1D cross-correlations with a template spectrum and with a template/model of the slit (\href{https://www.stsci.edu/files/live/sites/www/files/home/hst/instrumentation/stis/documentation/instrument-science-reports/_documents/199619.pdf}{Hulbert et al. 1996}; \href{https://www.stsci.edu/files/live/sites/www/files/home/hst/instrumentation/stis/documentation/instrument-science-reports/_documents/199812.pdf}{Hodge et al. 1998}; see \href{https://hst-docs.stsci.edu/stisdhb/chapter-3-stis-calibration/3-4-descriptions-of-calibration-steps#id-3.4DescriptionsofCalibrationSteps-3.4.233.4.23WAVECORR:WavecalCorrection}{Section 3.4.23 of the STIS Data Handbook}).
The zero-point offsets are recorded in the {\small SHIFTA1} (for x) and {\small SHIFTA2} (for y) keywords in the flt or crj image headers.
 
The exposure times for the default ``auto-wavecal'' exposures were originally determined from pre-launch lamp exposures (Lindler et al. 1996), with a rough scaling by (1/aperture area).\footnotemark
\footnotetext{Note that the actual wavecal exposure times are given in the image headers of the raw wav.fits files.  
The values given in the Astronomers' Proposal Tool (APT) orbit planner are padded somewhat (typically by about 10s).}
Unfortunately, however, all three of the Pt/Cr-Ne lamps have been fading -- at different rates, but particularly at the shortest wavelengths for all three (e.g., \href{https://www.stsci.edu/files/live/sites/www/files/home/hst/instrumentation/stis/documentation/instrument-science-reports/_documents/2011_01.pdf}{Pascucci et al. 2011}; \href{https://www.stsci.edu/files/live/sites/www/files/home/hst/instrumentation/stis/documentation/instrument-science-reports/_documents/2015_02.pdf}{Sonnentrucker 2015}; \href{https://www.stsci.edu/files/live/sites/www/files/home/hst/instrumentation/stis/documentation/instrument-science-reports/_documents/2017_04.pdf}{Peeples 2017}; \href{https://www.stsci.edu/files/live/sites/www/files/home/hst/instrumentation/stis/documentation/instrument-science-reports/_documents/2018_04.pdf}{Welty 2018}).
For example, the emission lines in LINE lamp spectra for the G140M/1218 setting decreased in strength by more than a factor of 30 between STIS installation (1997) and 2016, including during the period when STIS was not in operation (Figure~2 in \href{https://www.stsci.edu/files/live/sites/www/files/home/hst/instrumentation/stis/documentation/instrument-science-reports/_documents/2017_04.pdf}{Peeples 2017}; see also Appendix~A). 
Such fading was unexpected, as similar Pt/Cr-Ne lamps showed no significant reductions in line strength in ground-based accelerated aging tests (e.g., \href{https://ui.adsabs.harvard.edu/abs/2004SPIE.5488..679K/abstract}{Kerber et al. 2004}; \href{https://ui.adsabs.harvard.edu/abs/2006hstc.conf..318K/abstract}{Kerber et al. 2006}).

\begin{figure}[!h]
  \centering
  \includegraphics[width=170mm]{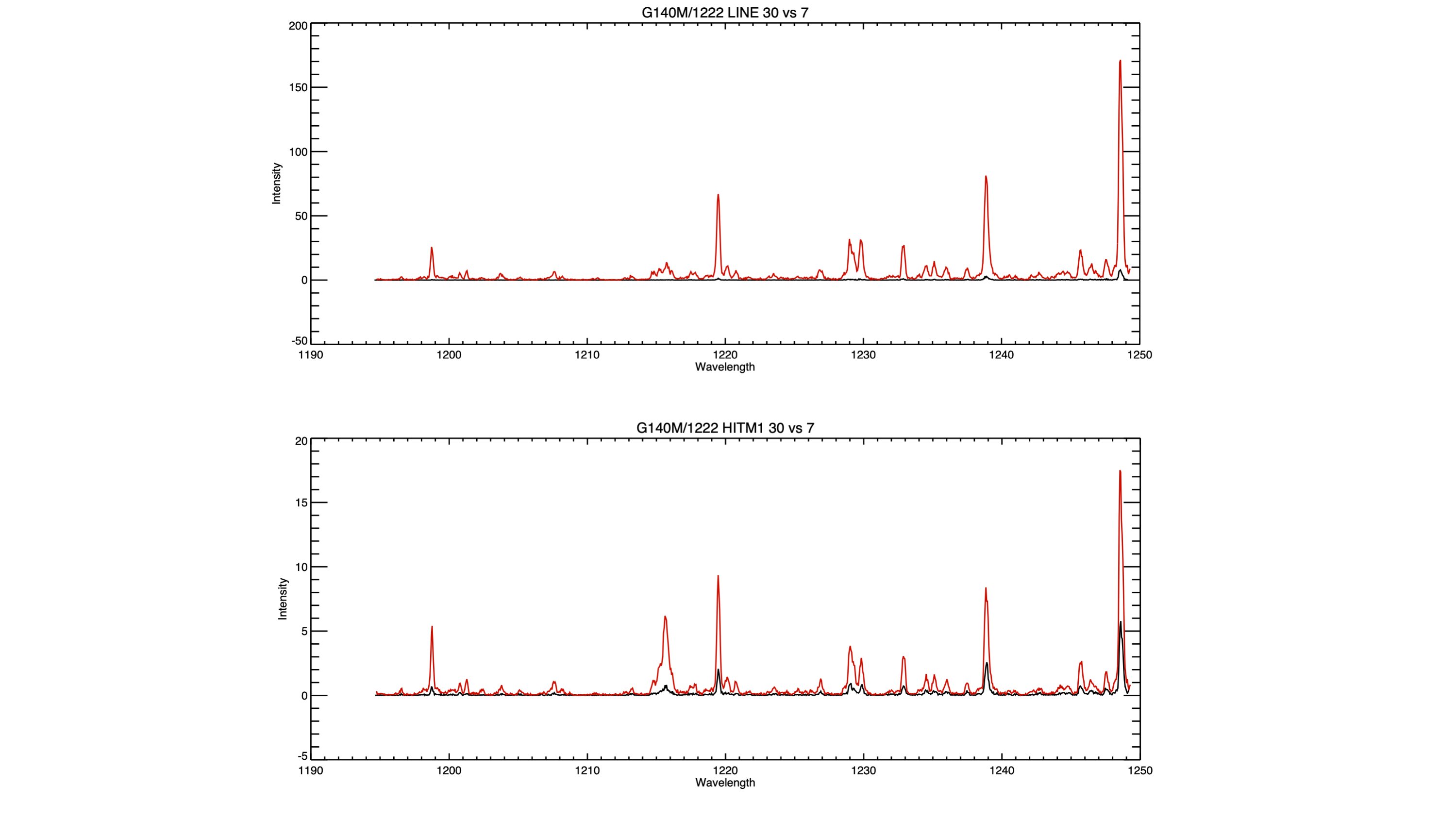}
    \caption{Pt/Cr-Ne wavecal spectra at the G140M/1222 setting, obtained with the LINE lamp (top) and the HITM1 lamp (bottom).
In each panel, the spectra in red were obtained in 1997/1998 (cycle 7), and the spectra in black were obtained in 2024 (cycle 30).
While the emission lines in this setting were initially much stronger in the LINE lamp spectrum, they also have faded more (especially at the shortest wavelengths) -- so that the lines below $\sim$1240 \AA\ are now stronger in the HITM1 spectrum.}
    \label{fig:gm1222}
\end{figure}

Figure~\ref{fig:gm1222} compares wavecals obtained between 1997/1998 (cycle 7) and 2024 (cycle 30) with both the LINE and HITM1 lamps, for the G140M/1222/52x0.1 setting.
Over that period, the lines near 1250 and 1200 \AA\ faded by factors of about 20 and more than 100 (respectively) for the LINE lamp and by factors of about 3 and 8 (respectively) for the HITM1 lamp.
Given that the lines from about 1200-1250 \AA\ were initially stronger by factors of 5-10 for the LINE lamp than for the HITM1 lamp, the net result is that the current HITM1 lamp line strengths are factors of about 40-30 weaker than the initial LINE lamp line strengths -- but comparable to or stronger than the current LINE lamp strengths -- over that wavelength range.
The lines in the less-used HITM2 lamp are a factor of 2-3 stronger than those in the HITM1 lamp, and appear to be declining at a similar rate.

\begin{figure}[!h]
  \centering
  \includegraphics[width=170mm]{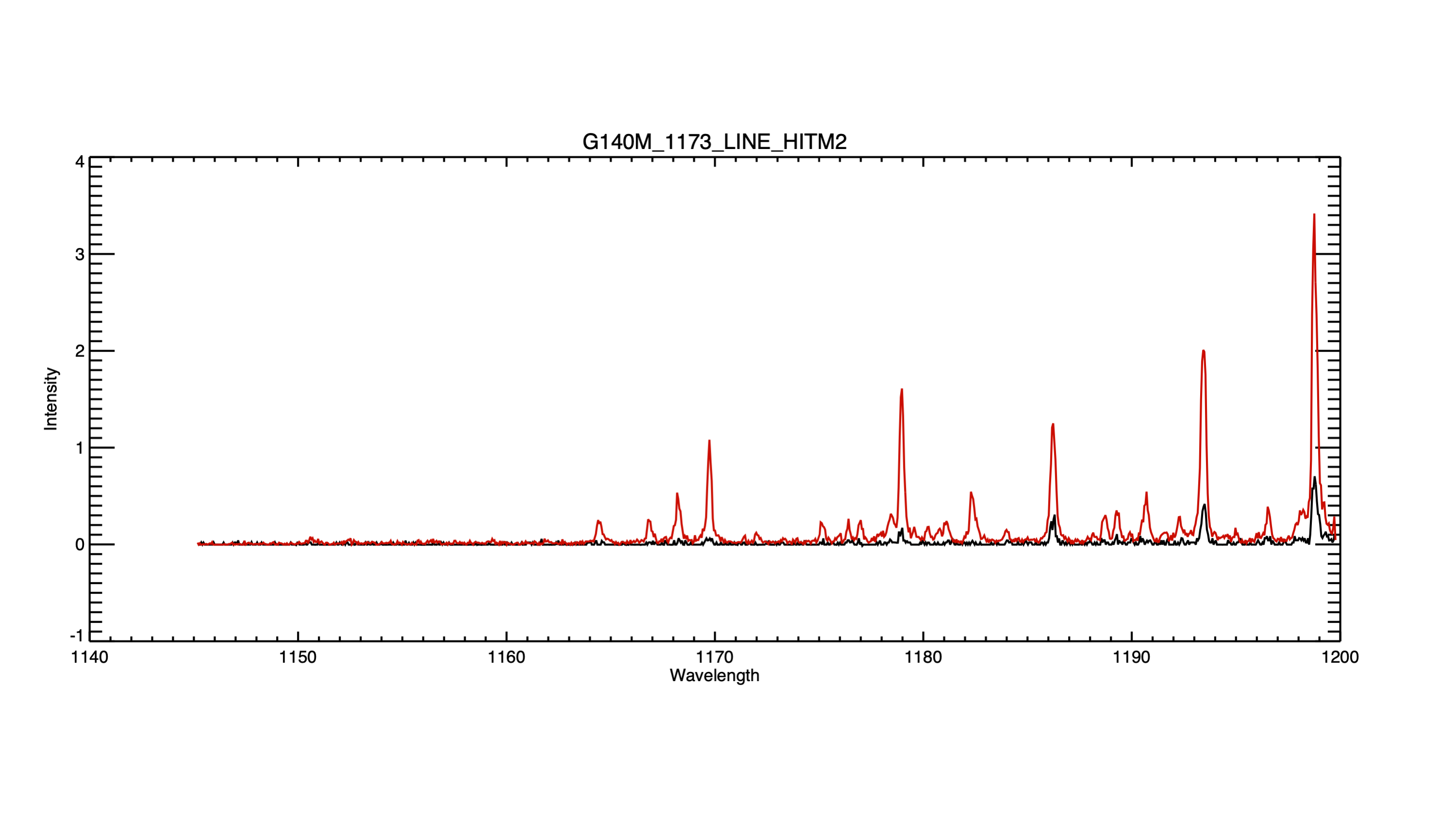}
    \caption{Pt/Cr-Ne lamp spectra at the G140M/1173 setting.
In black is a 46s LINE lamp spectrum from cycle 18 (PID 12414); in red is a 150s HITM2 lamp spectrum from cycle 19 (PID 12775).
Even then, the emission lines at these short wavelengths were stronger in the HITM2 spectrum -- and fading less rapidly -- which motivated a switch from LINE to HITM2 for the wavecal exposures in 2012.}
    \label{fig:spec}
\end{figure}

In an attempt to maintain robust determinations of the zero-point offsets for several of the shorter wavelength settings, the wavecal lamp was switched from LINE to HITM2 -- in 2012 for G140M/1173 and in 2016 for E140H/1234 and 1271 -- as by those dates the HITM2 lamp exhibited noticeably stronger lines at those short wavelengths (Figure~\ref{fig:spec}). 
No changes -- either in lamp or default exposure time -- have been made for the other settings to compensate for the fading of the lamps.
Observers using GO wavecals (instead of the default auto-wavecals), however, may specify longer exposure times (with justification and permission), in order to obtain more accurate wavelength zero points (\href{https://hst-docs.stsci.edu/stisihb/chapter-11-data-taking/11-2-exposure-sequences-and-contemporaneous-calibrations#id-11.2ExposureSequencesandContemporaneousCalibrations-Section11.2.111.2.1Auto-Wavecals}{STIS IHB, Sec.11.2.1}).

\begin{figure}[!h]
  \centering
  \includegraphics[width=170mm]{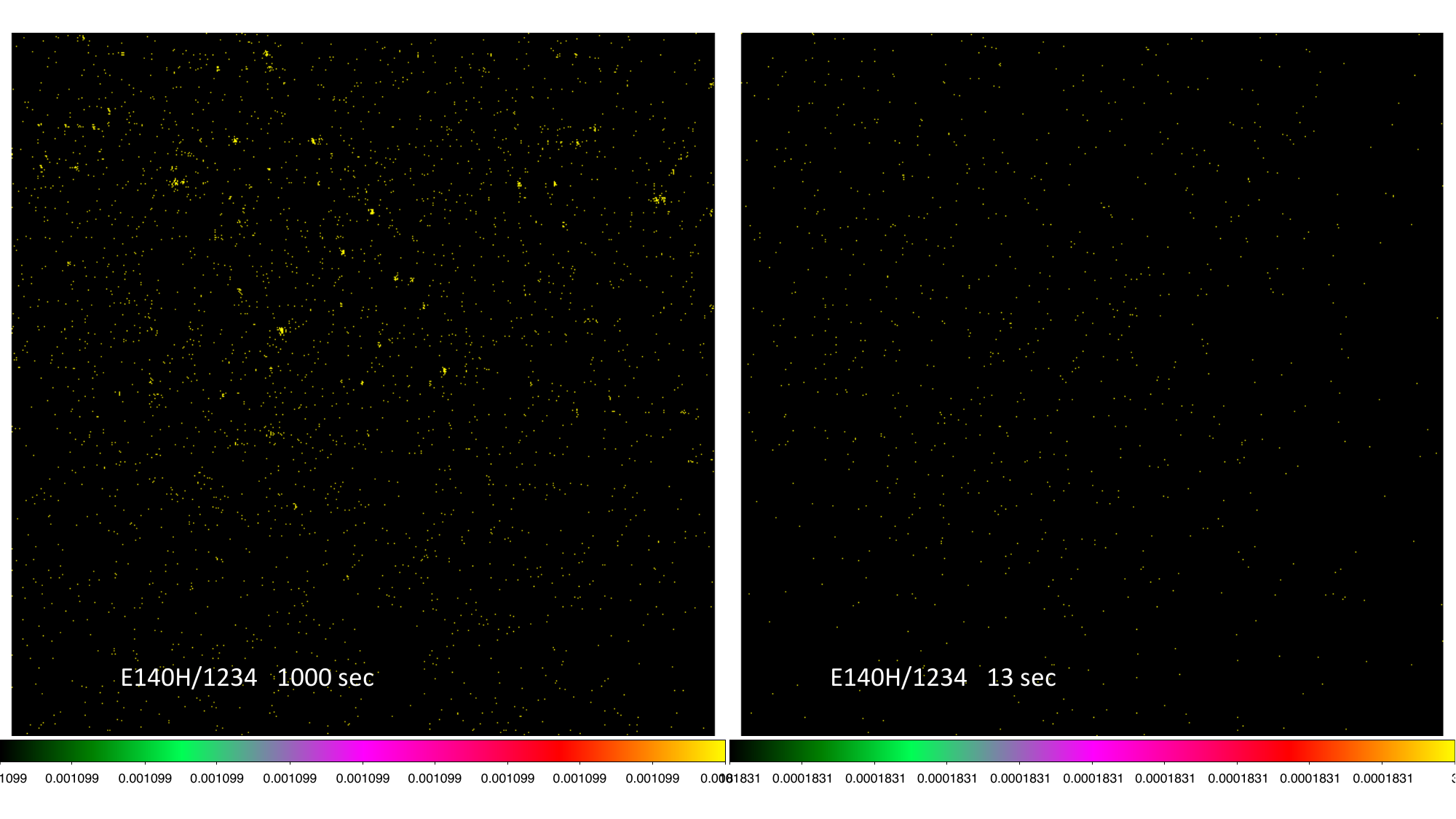}
    \caption{Pt/Cr-Ne lamp exposures at the E140H/1234 setting.
Individual echelle orders run approximately horizontally, with wavelengths increasing to the right within each order; the longer wavelength orders are at the top.
Many emission lines may be discerned, particularly in the upper half of the long (1000 sec) exposure (left panel), which was taken through the 0.2x0.09 aperture.
Those lines are very difficult to discern in the default (13 sec) wavecal exposure, taken through the 0.2x0.2 aperture, shown in the right-hand panel -- which nonetheless yielded fairly reasonable {\small SHIFTA} values.}
    \label{fig:lamp}
\end{figure}

Despite the significant fading of the lamps, the vast majority of wavecals still appear to yield fairly reasonable {\small SHIFTA} values -- which suggests that the initial default exposure times were quite generous, that the method for determining the offsets is fairly robust, and that the few changes that have been made have been effective.
However, there has been one obvious case of spurious offsets, for an observation with E140H/1234 taken in 2017 through the smallest 0.1x0.03 aperture (for dataset od9s02040), not long after the lamp was switched to HITM2 for that setting.
Moreover, the emission lines in the default wavecal exposures for some of the shorter wavelength E140H settings are now rather difficult to discern (e.g., Figure~\ref{fig:lamp}).
Finally, as described below, there is some scatter in the offsets obtained from the default wavecals.
It therefore seemed worthwhile to consider whether further adjustments to the default lamps and/or exposure times might be warranted, in order to ensure accurate wavelength zero points for STIS spectroscopic exposures.
In this ISR, we first investigate the behavior of the default wavecals for various echelle and first-order MAMA settings, by examining the trends in the {\small SHIFTA} values derived from those wavecals.
We then describe the results of simulations aimed at estimating the exposure times that would currently be needed for accurate determination of the wavelength zero-points for the shortest wavelength settings, using sub-samples of a long time-tagged E140H/1234 lamp exposure.
Two appendices provide further information regarding the fading of the lamps and the usage of the lamps.
Based on these investigations, we recommend that the default exposure times for several of the shortest wavelength E140H and G140M settings be increased by factors of 1.5-4.
Changing the lamp to HITM2 for several of those settings could reduce the exposure times needed, by a factor $\sim$2.

%

\begin{deluxetable}{cll}
    \tabcolsep 10pt
    \tablewidth{0pt}
    \tablecaption{STIS Wavecal Apertures \label{tab:aper}}
    \tabletypesize{\footnotesize}
    \tablehead{ \colhead{Grating} & \colhead{Use Same Aperture} & 
                \colhead{All Others} }
    \startdata
E140M  & 0.1x0.03, 0.2x0.06, 0.2x0.2, 6x0.2                    & 0.2x0.06 \\
E140H  & 0.1x0.03, 0.2x0.09, 0.2x0.2, 6x0.2                    & 0.2x0.09 \\
E230M  & 0.1x0.03, 0.2x0.06, 0.2x0.2, 6x0.2                    & 0.2x0.06 \\
E230H  & 0.1x0.03, 0.1x0.09, 0.1x0.2, 0.2x0.09, 0.2x0.2, 6x0.2 & 0.2x0.09 \\
G140L  & 52x0.05, 52x0.1, 52x0.2                               & 52x0.05 \\
G140M  & 52x0.05, 52x0.1, 52x0.2                               & 52x0.1 \\
G230L  & 52x0.05, 52x0.1, 52x0.2, 31x0.05NDC                   & 31x0.05NDC \\
G230M  & 52x0.05, 52x0.1, 52x0.2                               & 52x0.1 \\
G230LB & 52x0.05, 52x0.1, 52x0.2                               & 52x0.1 \\
G230MB & 52x0.05, 52x0.1, 52x0.2                               & 52x0.1 \\
G430L  & 52x0.05, 52x0.1, 52x0.2                               & 52x0.1 \\
G430M  & 52x0.05, 52x0.1, 52x0.2                               & 52x0.1 \\
G750L  & 52x0.05, 52x0.1, 52x0.2                               & 52x0.1 \\
G750M  & 52x0.05, 52x0.1, 52x0.2                               & 52x0.1 \\
    \enddata
\end{deluxetable}

\clearpage
\ssection{Statistics of {\small SHIFTA} Values}\label{sec:stats}

The position of a STIS spectrum on the MAMA or CCD detectors depends on the location of the nominal center of the aperture used, any deliberate offsets applied (e.g., the repeller wire and monthly offsets used for MAMA spectra, any POS-TARGs or dither/pattern specifications), and the generally small (and not entirely predictable) offsets characterizing the specific wavelength setting.
The contemporaneous wavecals are designed to measure the last of those factors. 
Examination of the {\small SHIFTA} values obtained from those wavecals, for the many STIS spectra in the MAST archive, can provide some information regarding the mean (x,y) offsets and typical scatter for each setting -- and whether the reliability of any of those values might be degrading as the lamps have faded.

\begin{figure}[!h]
  \centering
  \includegraphics[width=170mm]{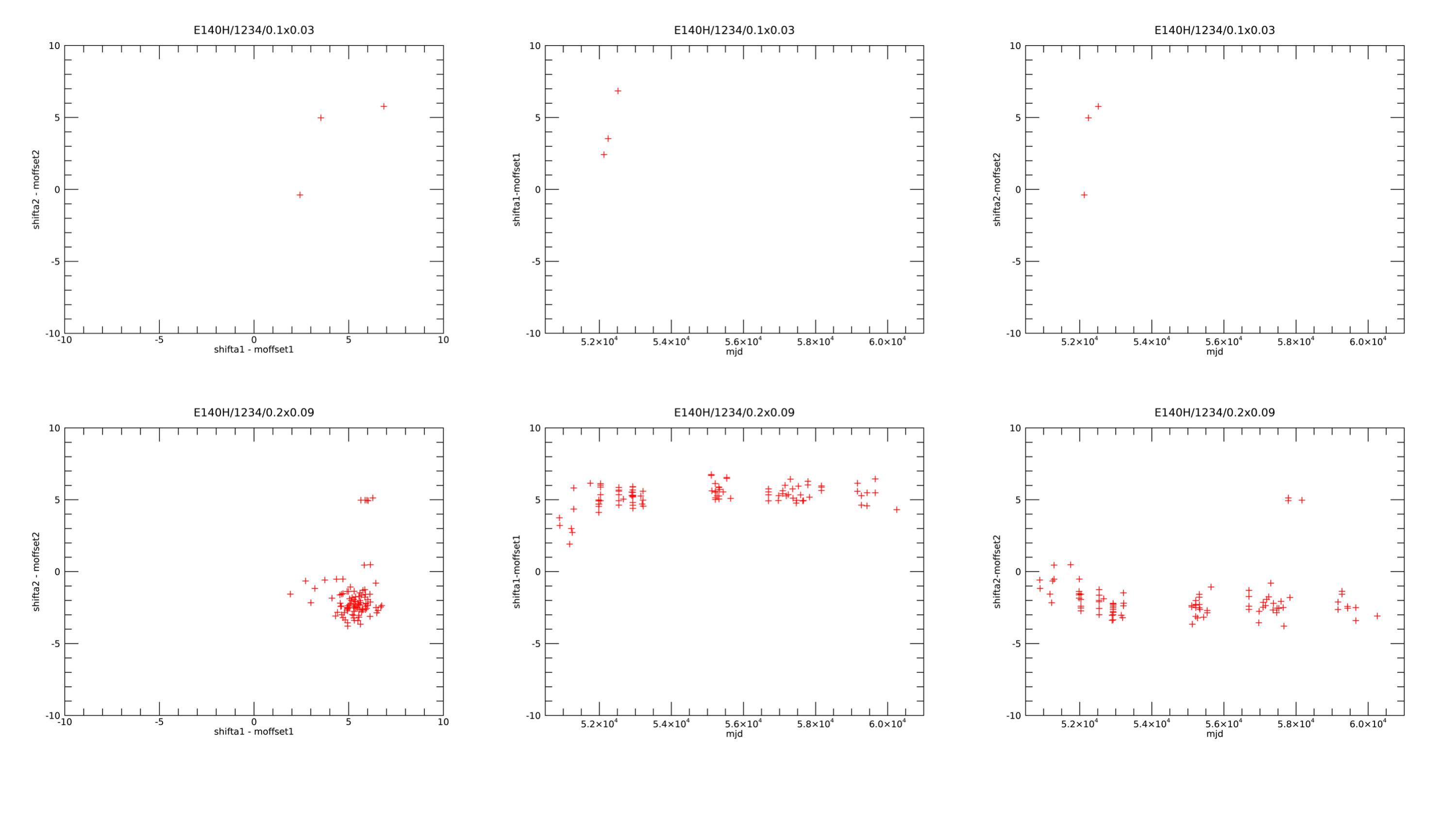}
    \caption{{\small SHIFTA $-$ MOFFSET} values for E140H/1234 observations taken through the 0.1x0.03 and 0.2x0.09 apertures (upper and lower panels).
The left-hand panels show {\small SHIFTA2} $-$ {\small MOFFSET2} vs. {\small SHIFTA1} $-$ {\small MOFFSET1}; the center and right-hand panels show {\small SHIFTA1} $-$ {\small MOFFSET1} and {\small SHIFTA2} $-$ {\small MOFFSET2} vs. MJD.
In the left-hand panels, most of the points fall within a main clump near (5, -2), but there can be one or more outlying sub-clumps and/or more isolated ``outliers'' as well.
The sub-clumps appear to reflect some combination of increased scatter during the period when monthly offsets were applied to the echelle settings (MJD $\sim$ 50825--52500), changes in the slit wheel and/or mode select mechanism values adopted for each setting, and/or some non-reproducibility in the results of those values.}
    \label{fig:stats1}
\end{figure}

\begin{figure}[!h]
  \centering
  \includegraphics[width=170mm]{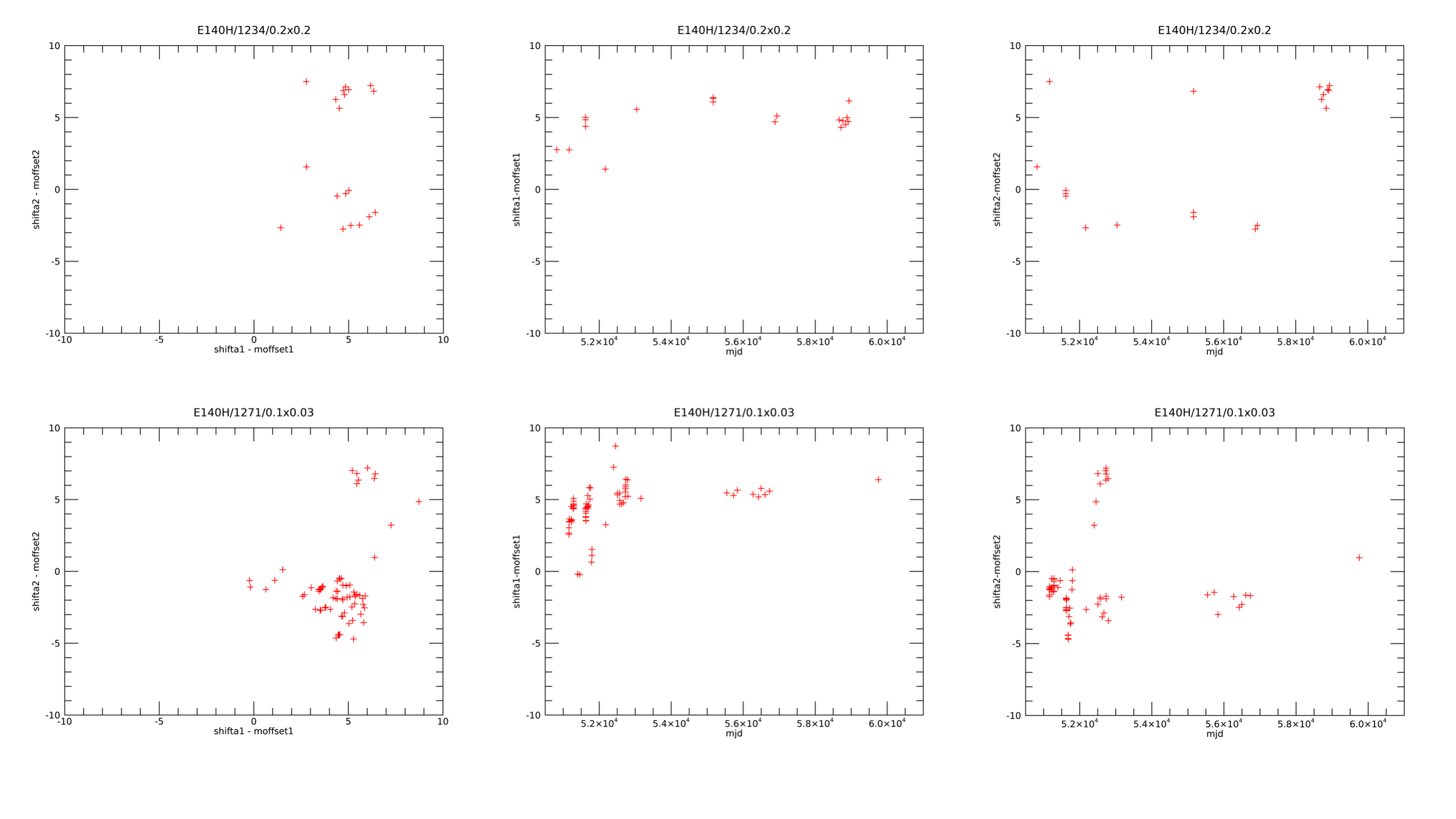}
  \includegraphics[width=170mm]{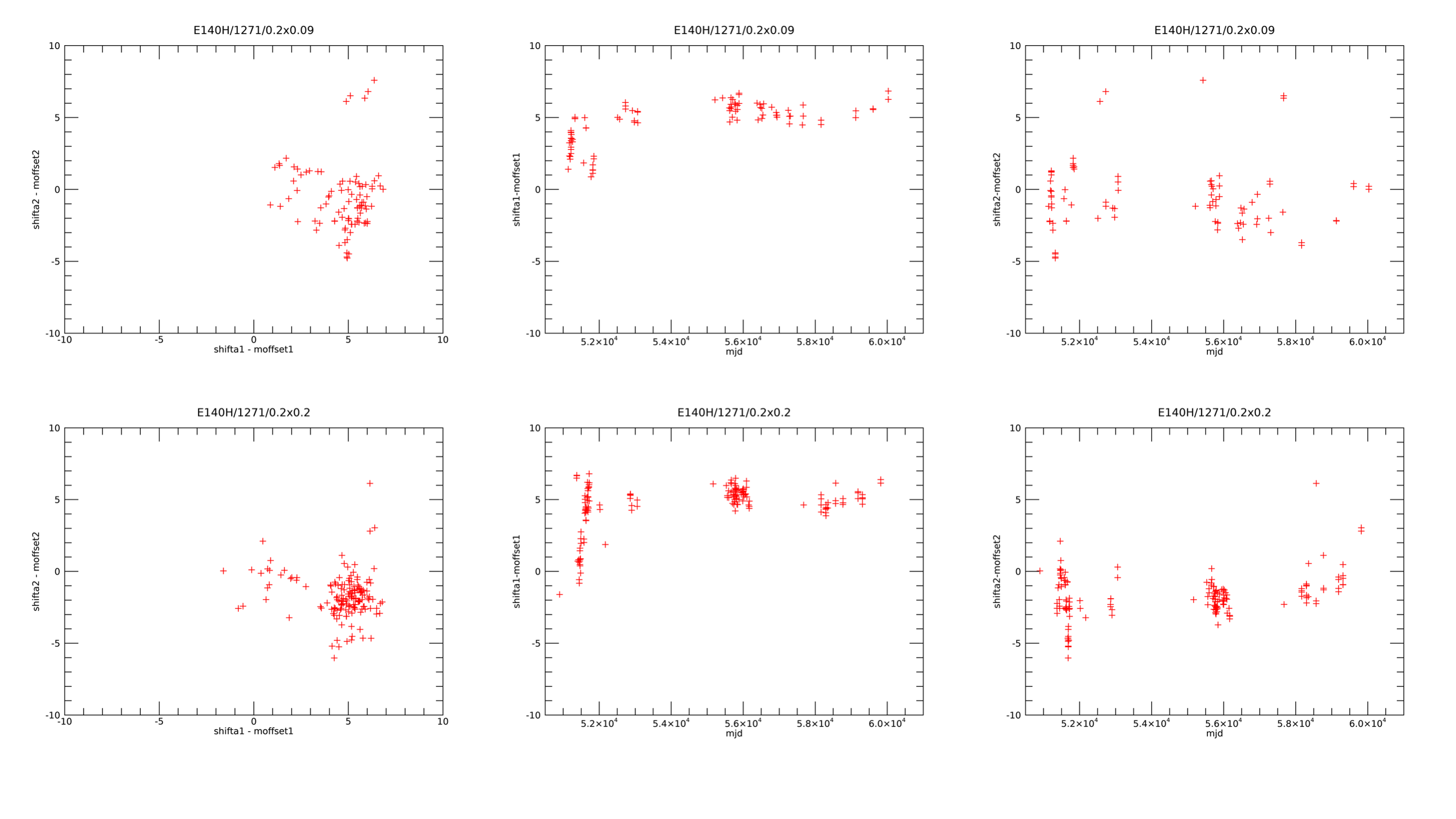}
    \caption{{\small SHIFTA $-$ MOFFSET} values (as in Fig.~\ref{fig:stats1}) for E140H/1234 ond 1271 observations taken through several apertures.  
The most recent points for E140H/1234/0.2x0.2 are all for eta Car, which is often discrepant in {\small SHIFTA2$-$MOFFSET2}.}
    \label{fig:stats2}
\end{figure}

\begin{figure}[!h]
  \centering
  \includegraphics[width=170mm]{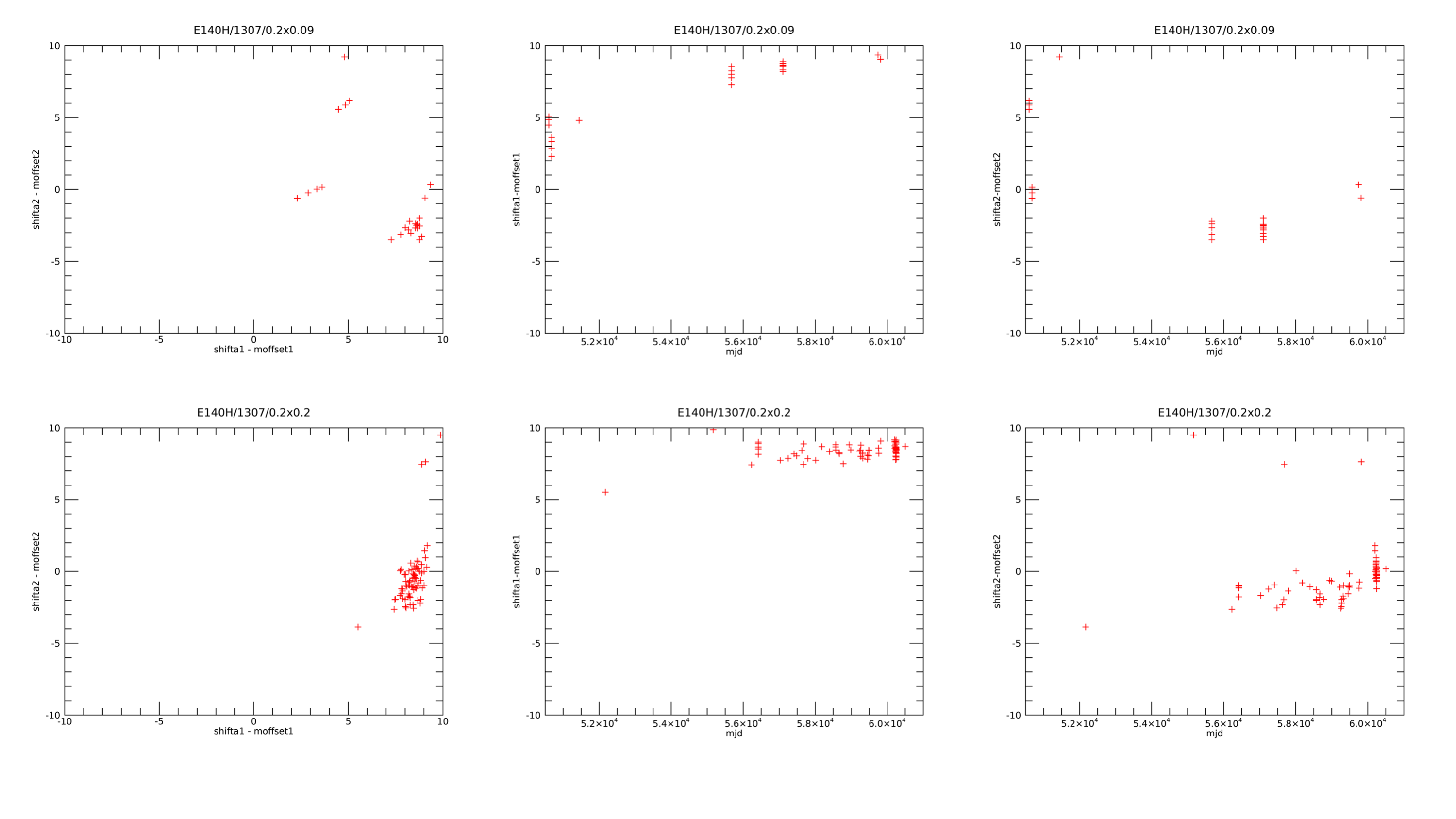}
  \includegraphics[width=170mm]{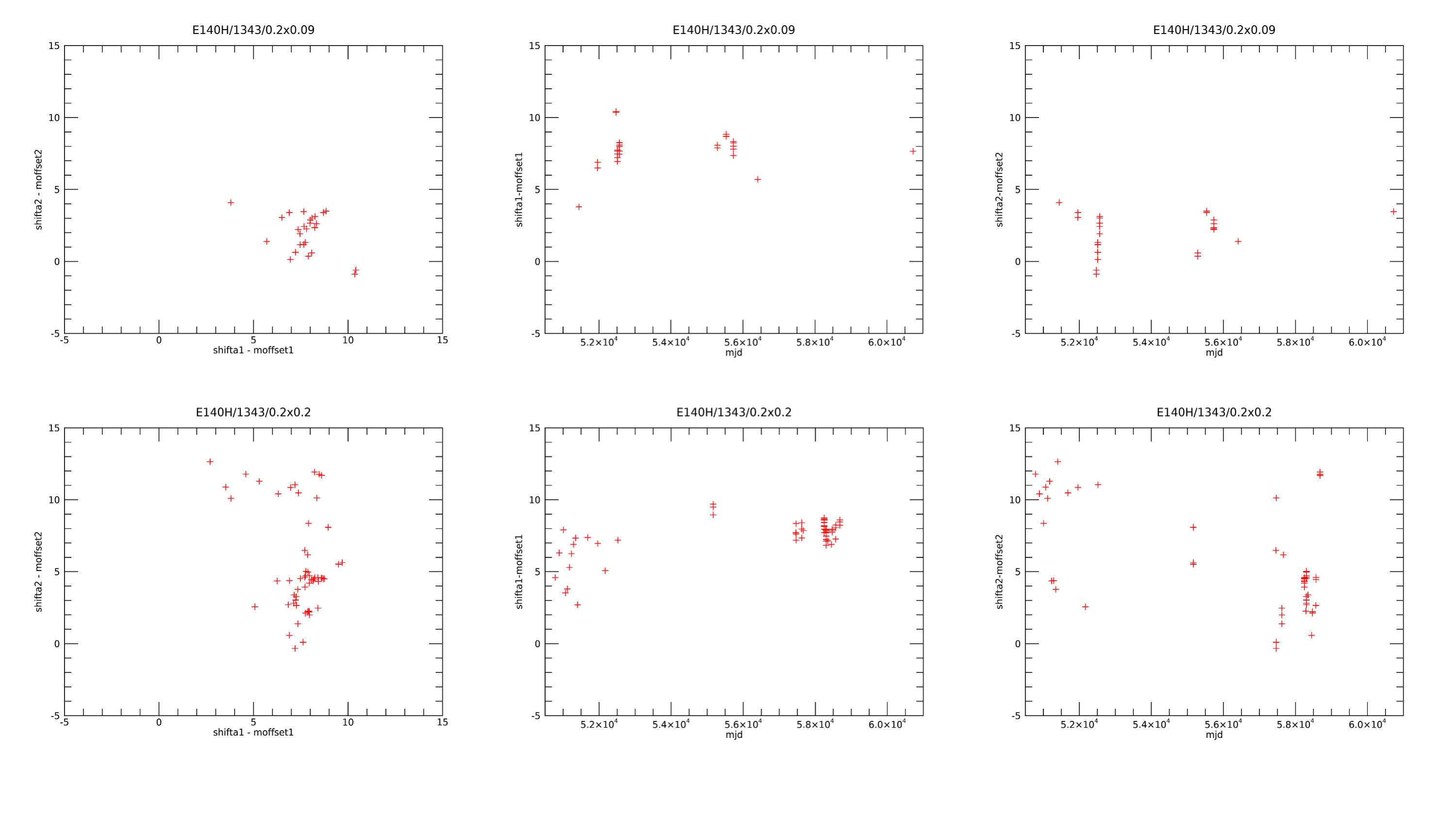}
    \caption{{\small SHIFTA $-$ MOFFSET} values (as in Fig.~\ref{fig:stats1}) for E140H/1307 and 1343 observations taken through the 0.2x0.09 and 0.2x0.2 apertures.}
    \label{fig:stats3}
\end{figure}

\begin{figure}[!h]
  \centering
  \includegraphics[width=170mm]{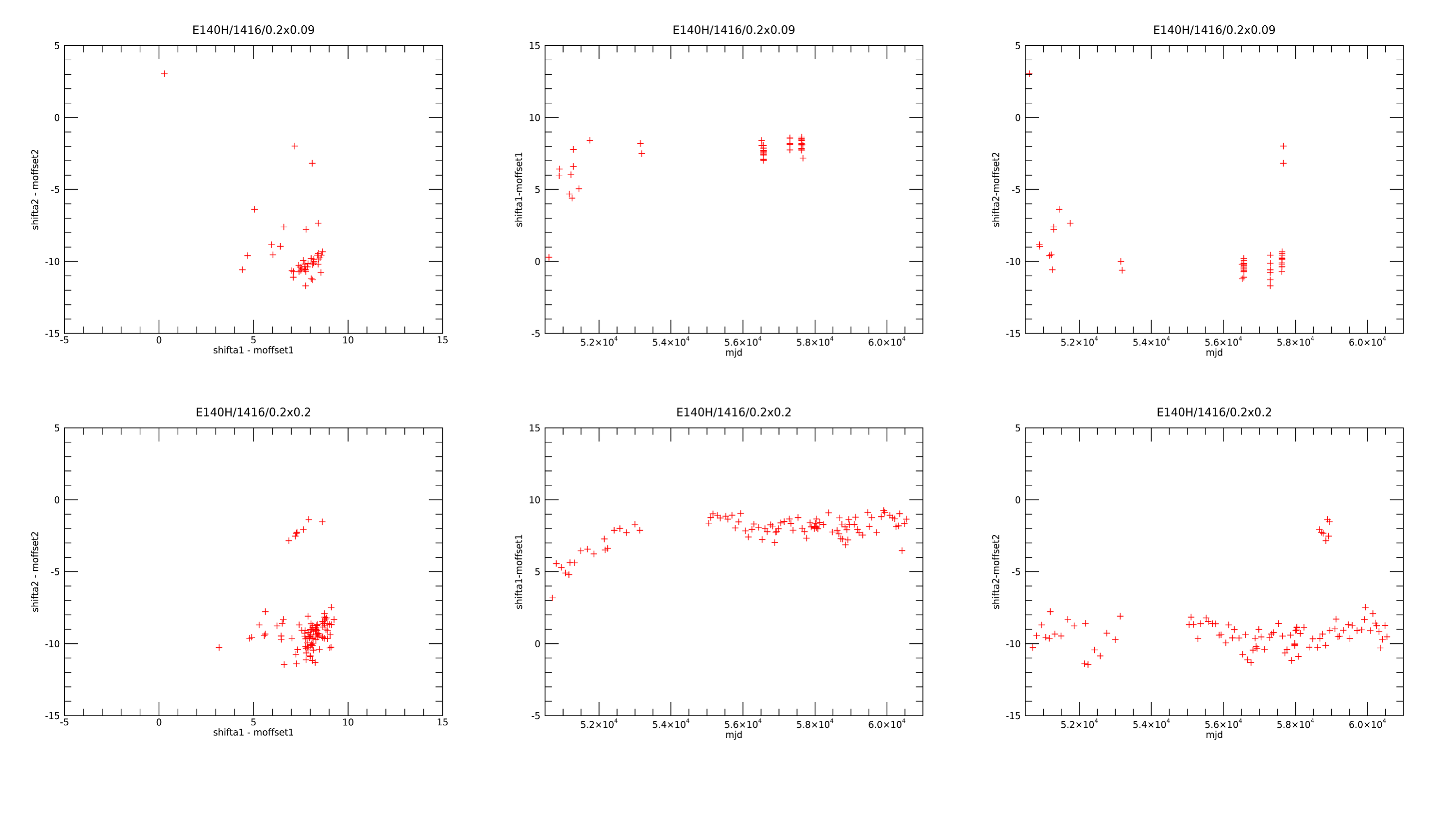}
  \includegraphics[width=170mm]{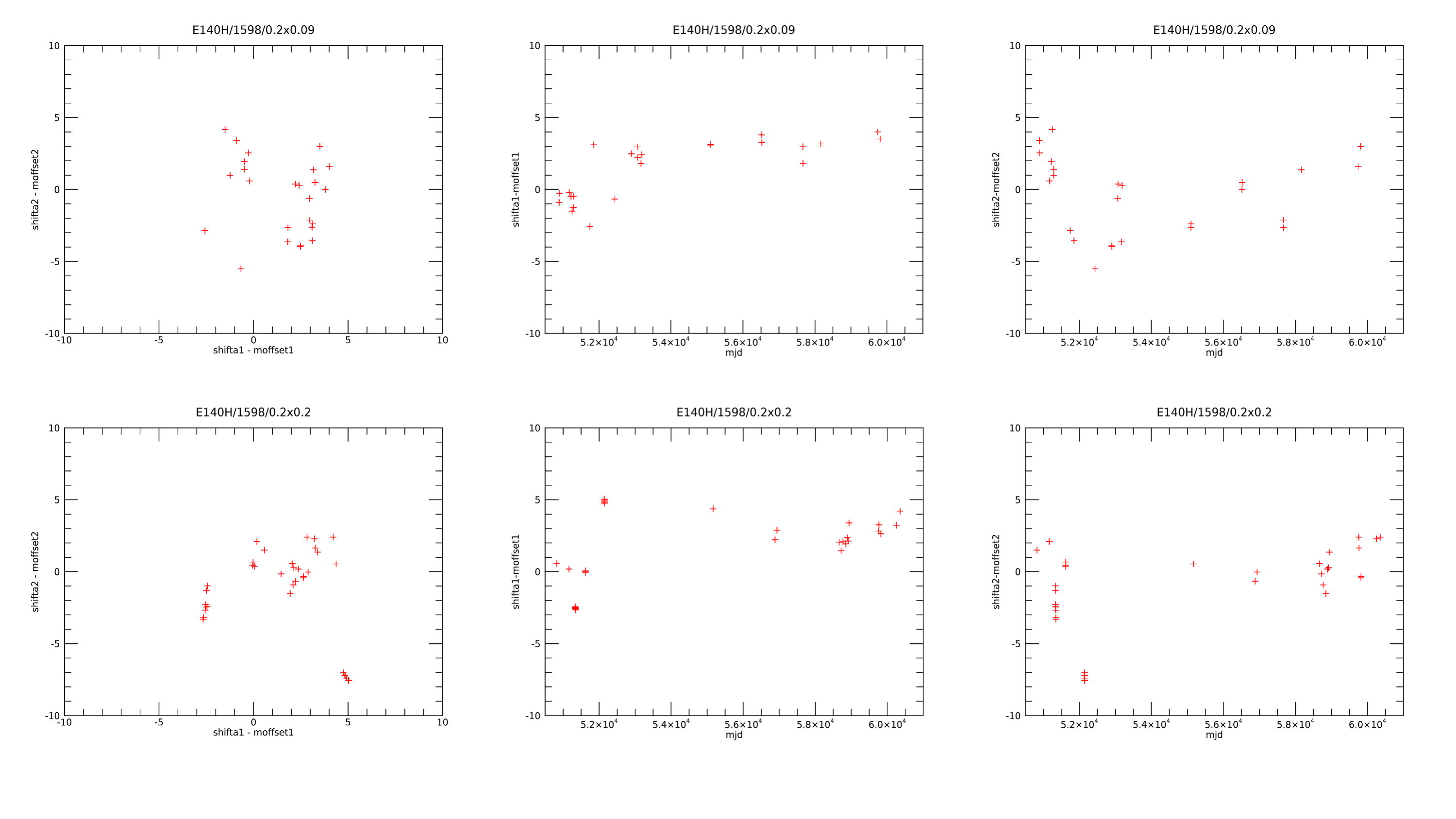}
    \caption{{\small SHIFTA $-$ MOFFSET} values (as in Fig.~\ref{fig:stats1}) for E140H/1416 and 1598 observations taken through the 0.2x0.09 and 0.2x0.2 apertures.
The recent high {\small SHIFTA2$-$MOFFSET2} points for E140H/1416/0.2x0.2 are for eta Car.}
    \label{fig:stats4}
\end{figure}

\begin{figure}[!h]
  \centering
  \includegraphics[width=170mm]{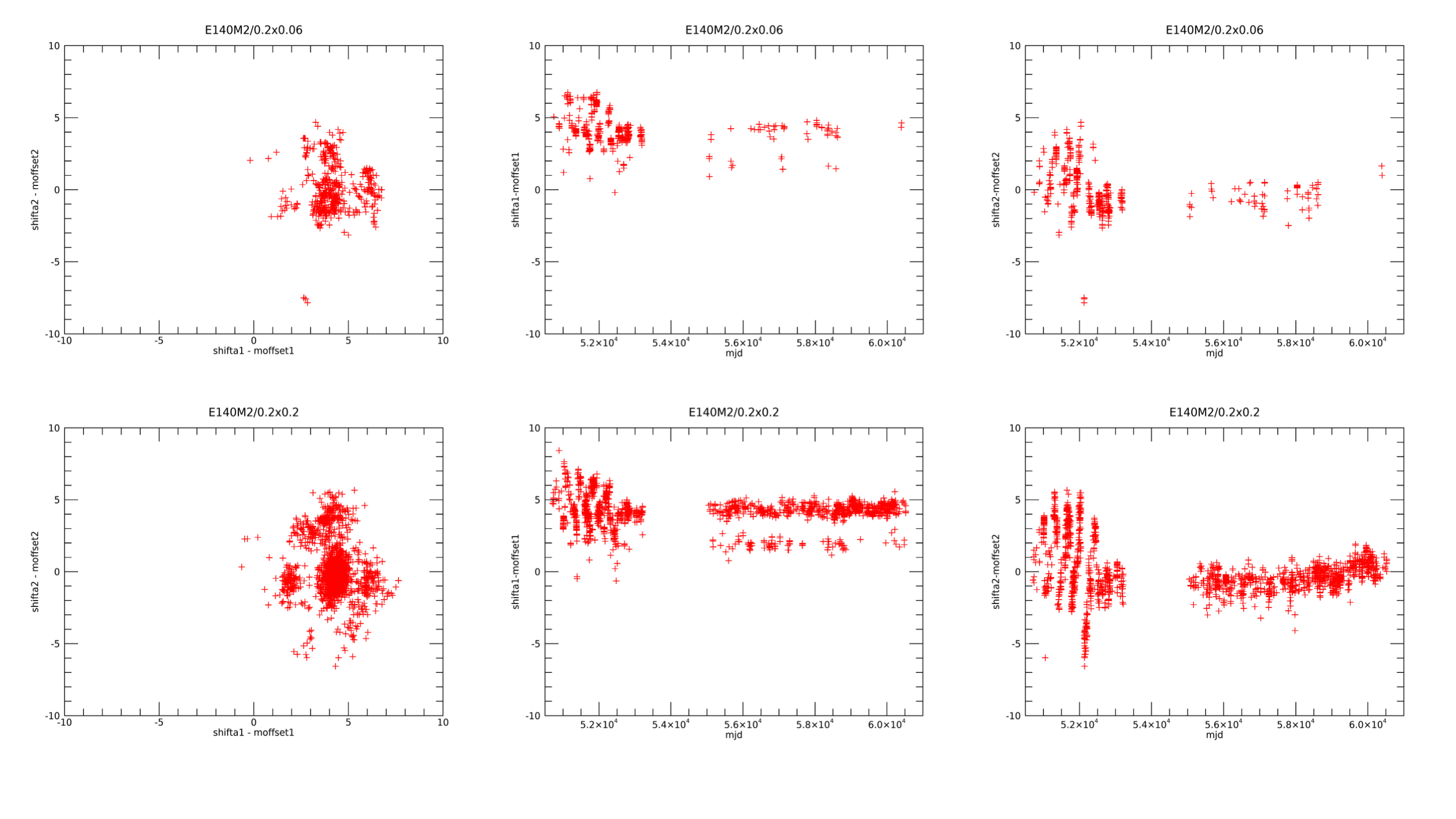}
    \caption{{\small SHIFTA $-$ MOFFSET} values (as in Fig.~\ref{fig:stats1}) for E140M/1425 observations taken through the 0.2x0.06 and 0.2x0.2 apertures.
Note the apparent bi-modality of {\small SHIFTA1 $-$ MOFFSET1} post-SM4.}
    \label{fig:stats5}
\end{figure}

\begin{figure}[!h]
  \centering
  \includegraphics[width=170mm]{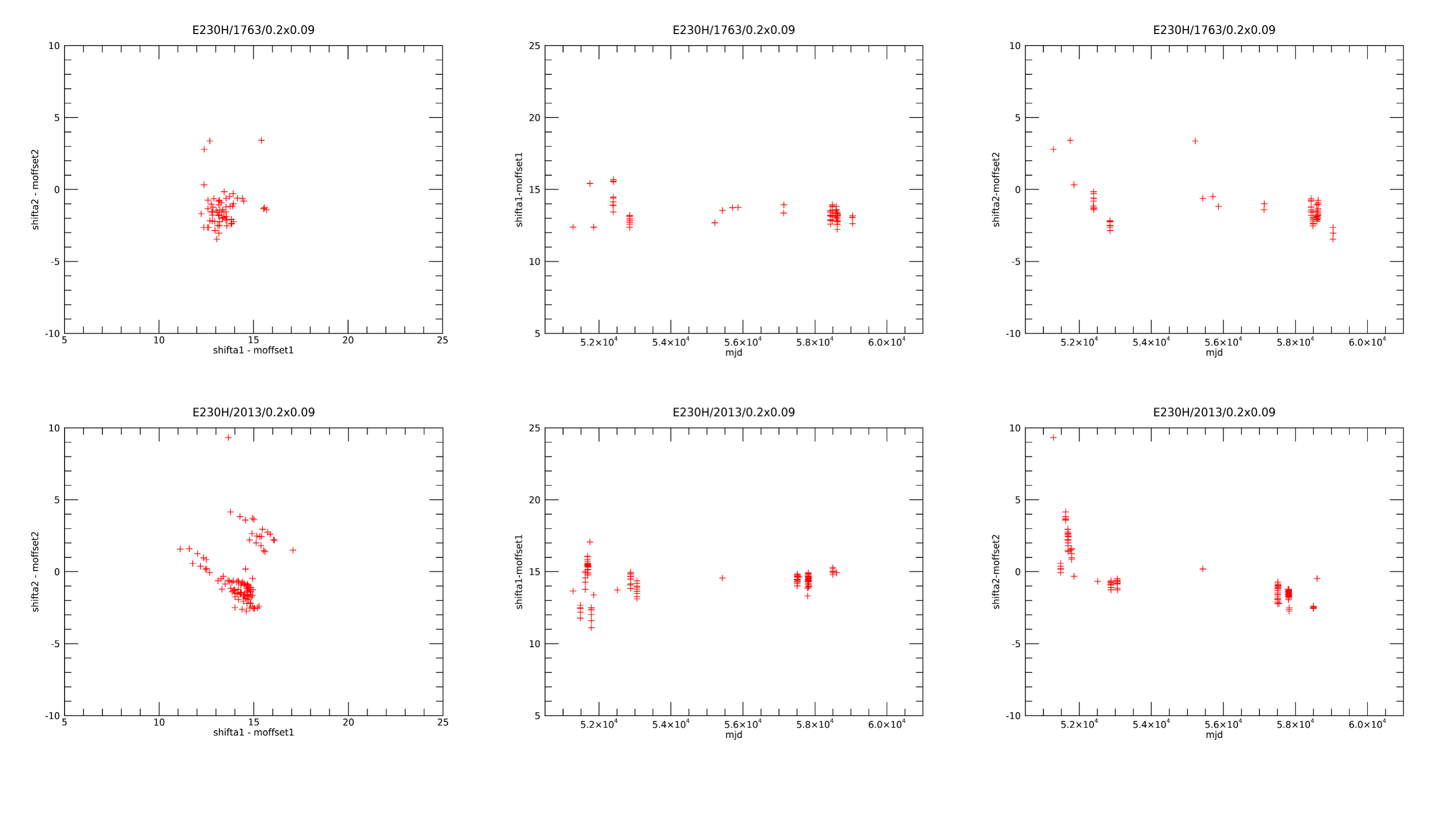}
  \includegraphics[width=170mm]{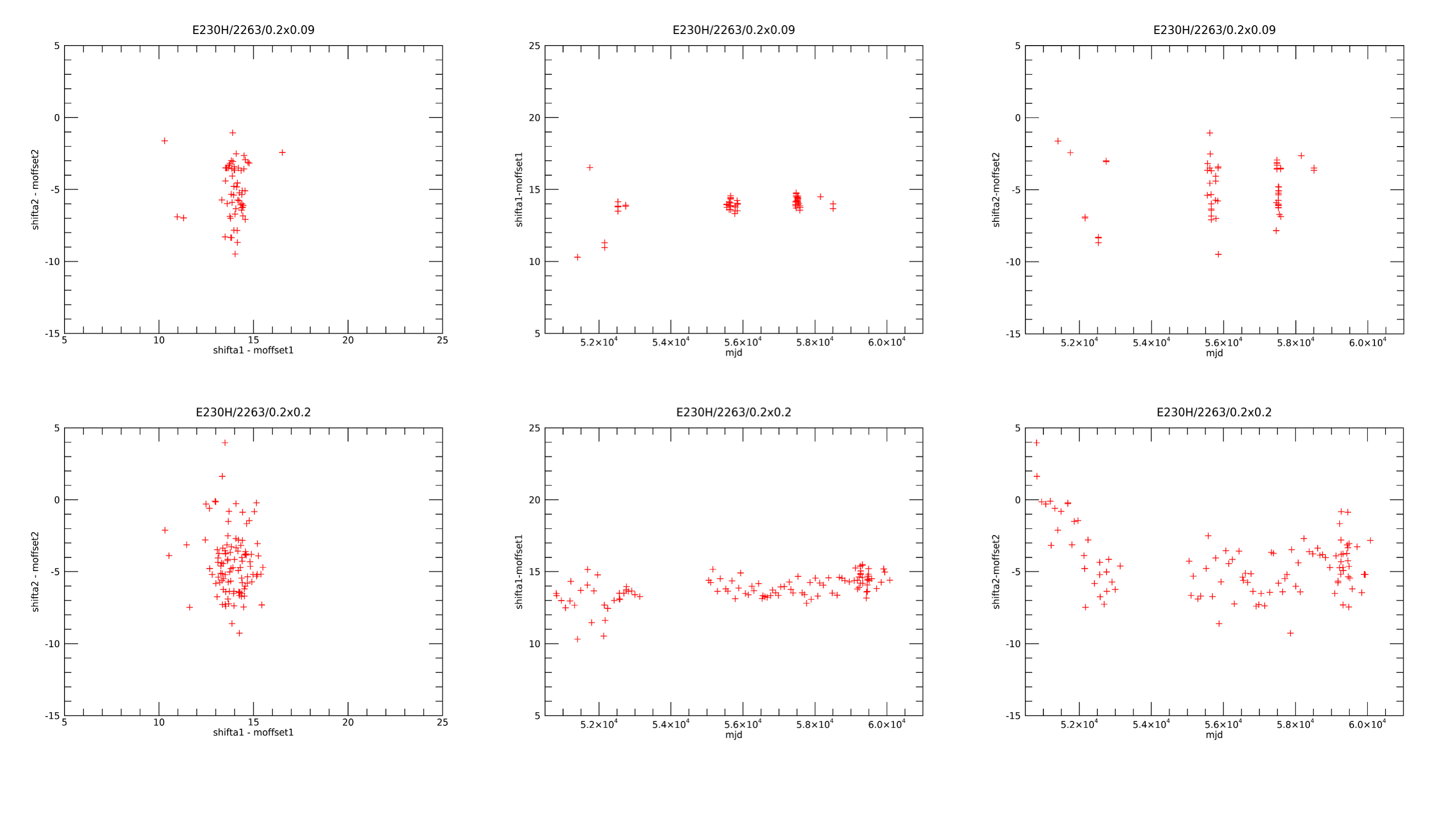}
    \caption{{\small SHIFTA $-$ MOFFSET} values (as in Fig.~\ref{fig:stats1}) for E230H/1763, 2013, and 2263 observations taken through the 0.2x0.09 and/or 0.2x0.2 apertures.}
    \label{fig:stats6}
\end{figure}

\begin{figure}[!h]
  \centering
  \includegraphics[width=170mm]{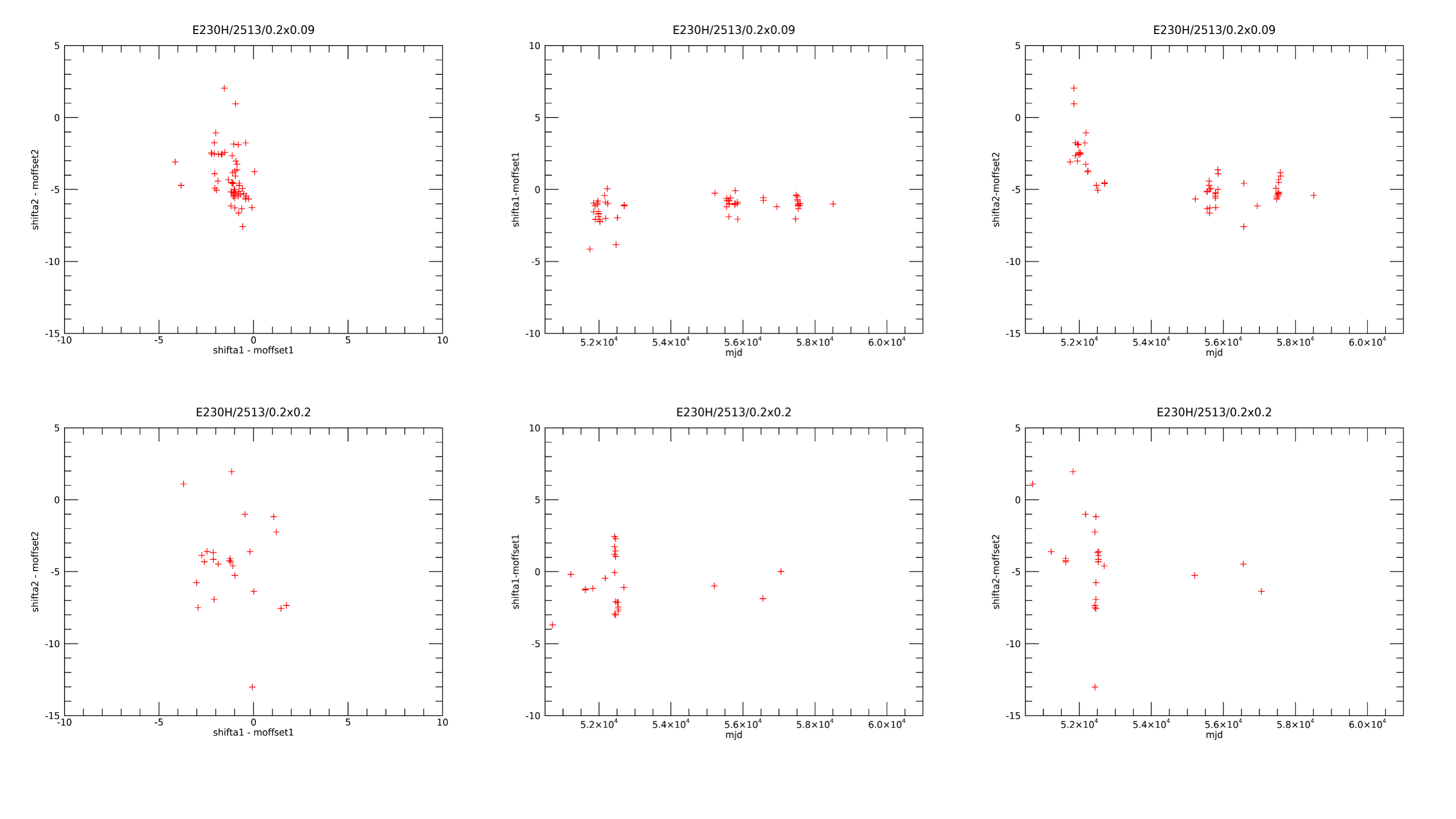}
  \includegraphics[width=170mm]{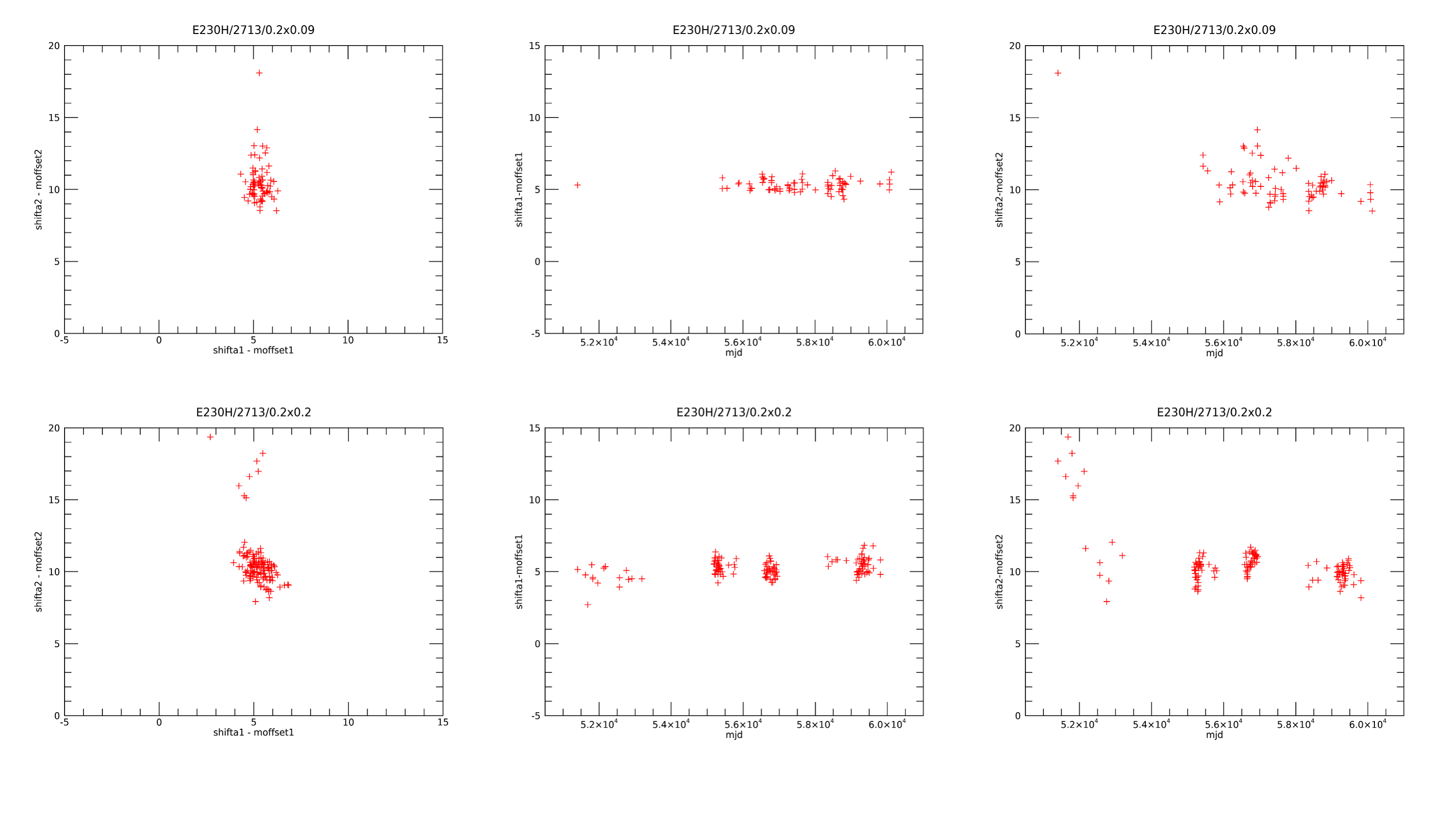}
    \caption{{\small SHIFTA $-$ MOFFSET} values (as in Fig.~\ref{fig:stats1}) for E230H/2513 and 2713 observations taken through the 0.2x0.09 and 0.2x0.2 apertures.}
    \label{fig:stats7}
\end{figure}

\begin{figure}[!h]
  \centering
  \includegraphics[width=170mm]{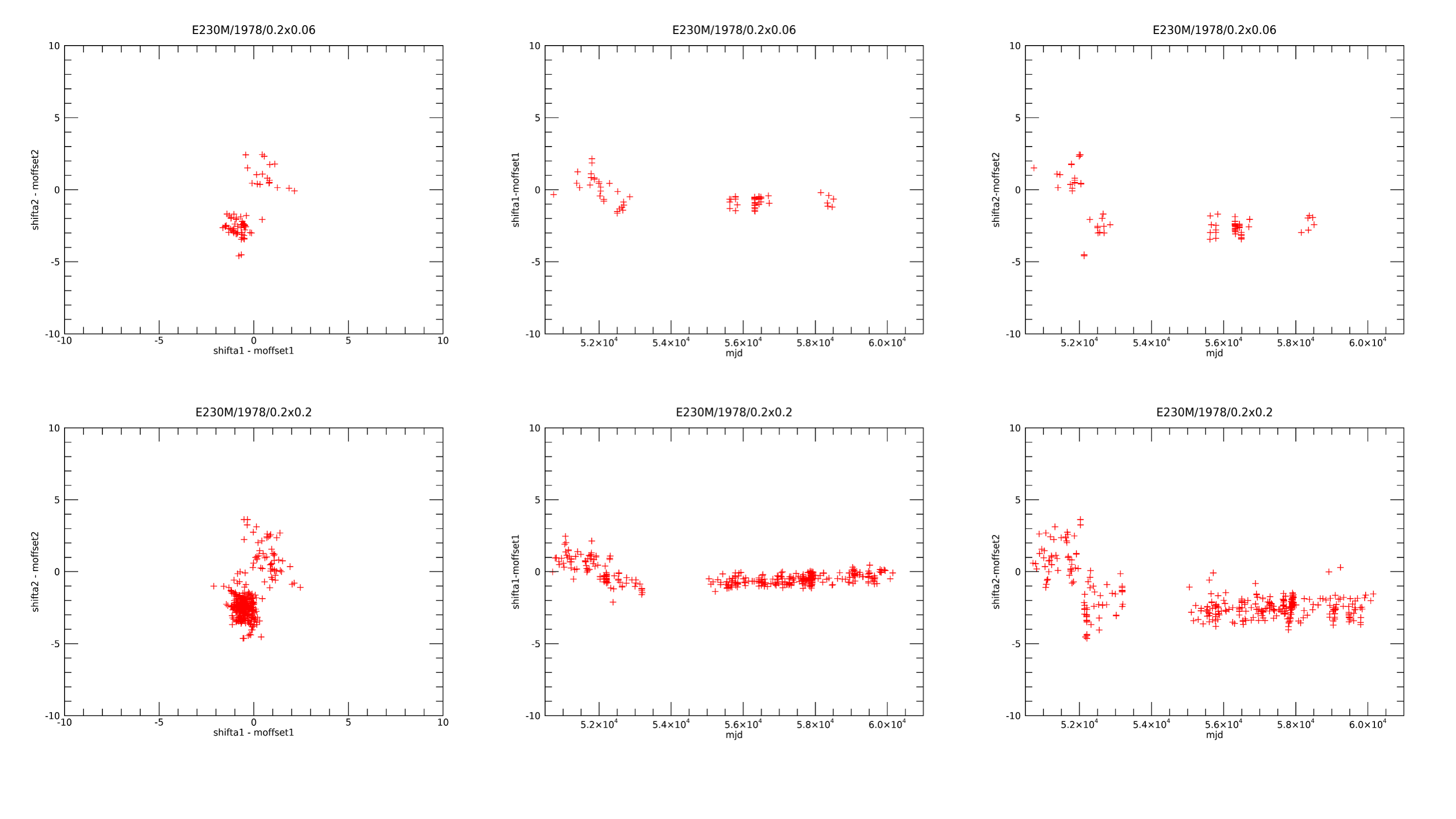}
  \includegraphics[width=170mm]{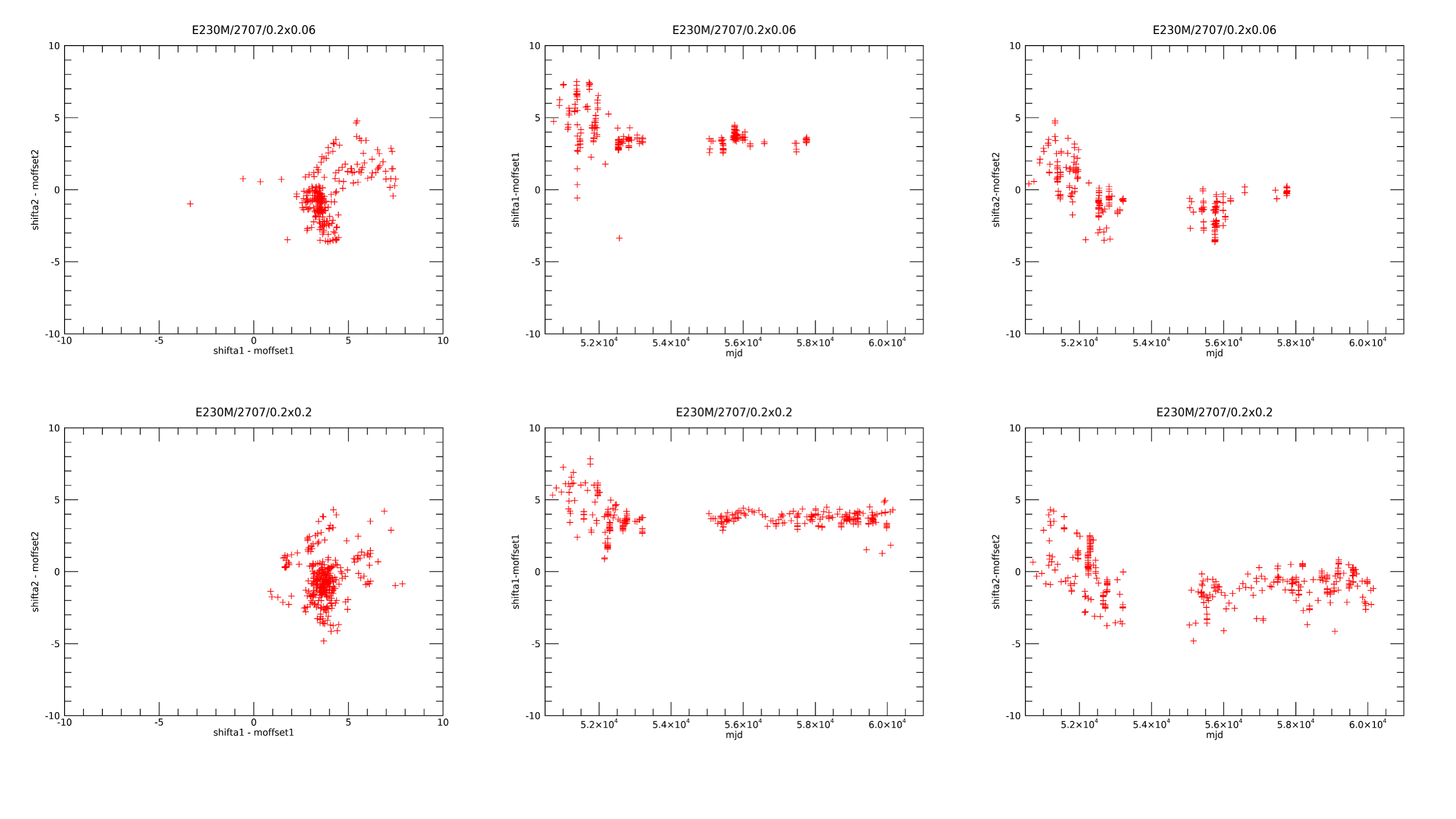}
    \caption{{\small SHIFTA $-$ MOFFSET} values (as in Fig.~\ref{fig:stats1}) for E230M/1978 and 2707 observations taken through the 0.2x0.06 and 0.2x0.2 apertures.}
    \label{fig:stats8}
\end{figure}

\begin{figure}[!h]
  \centering
  \includegraphics[width=170mm]{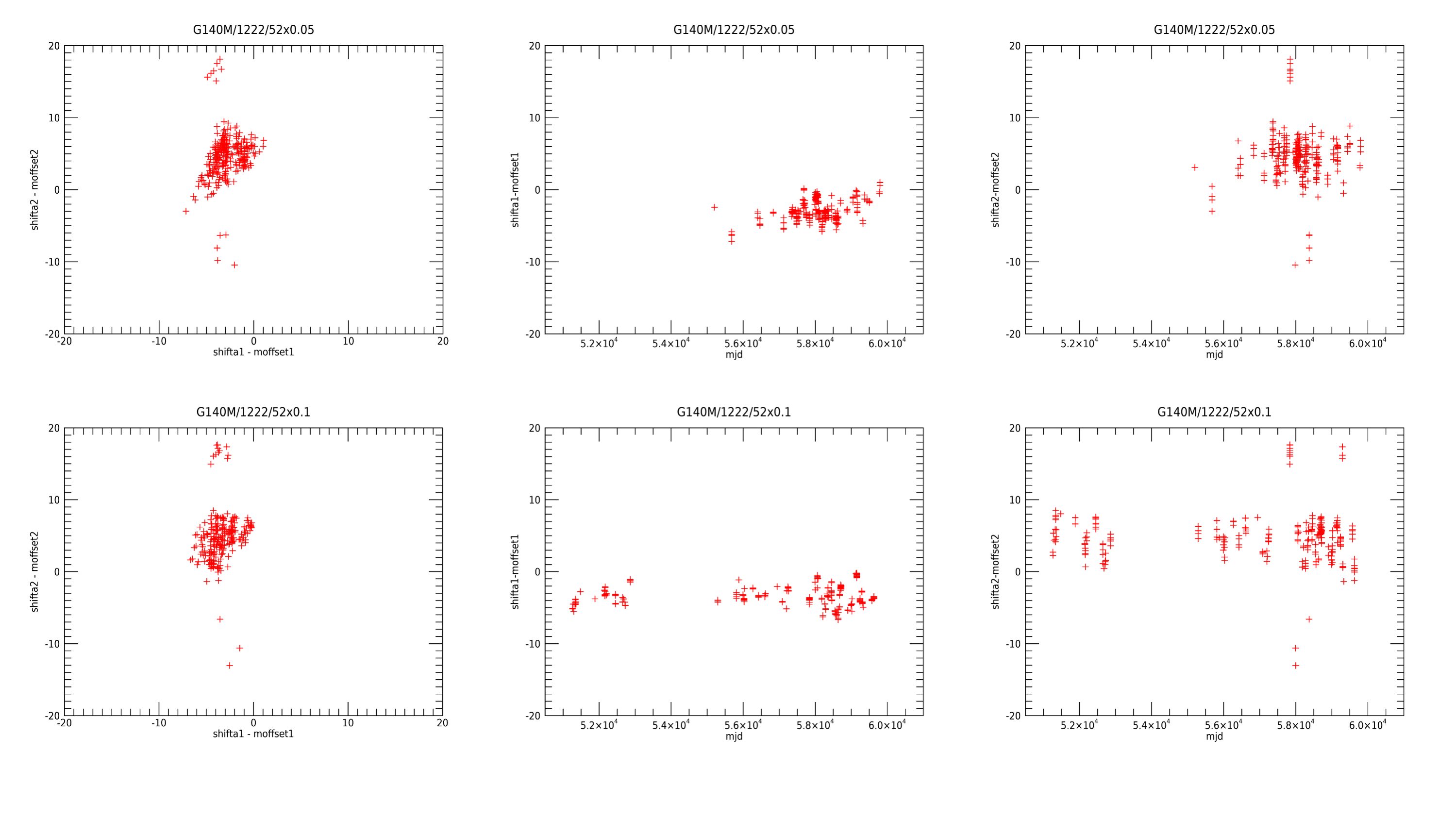}
  \includegraphics[width=170mm]{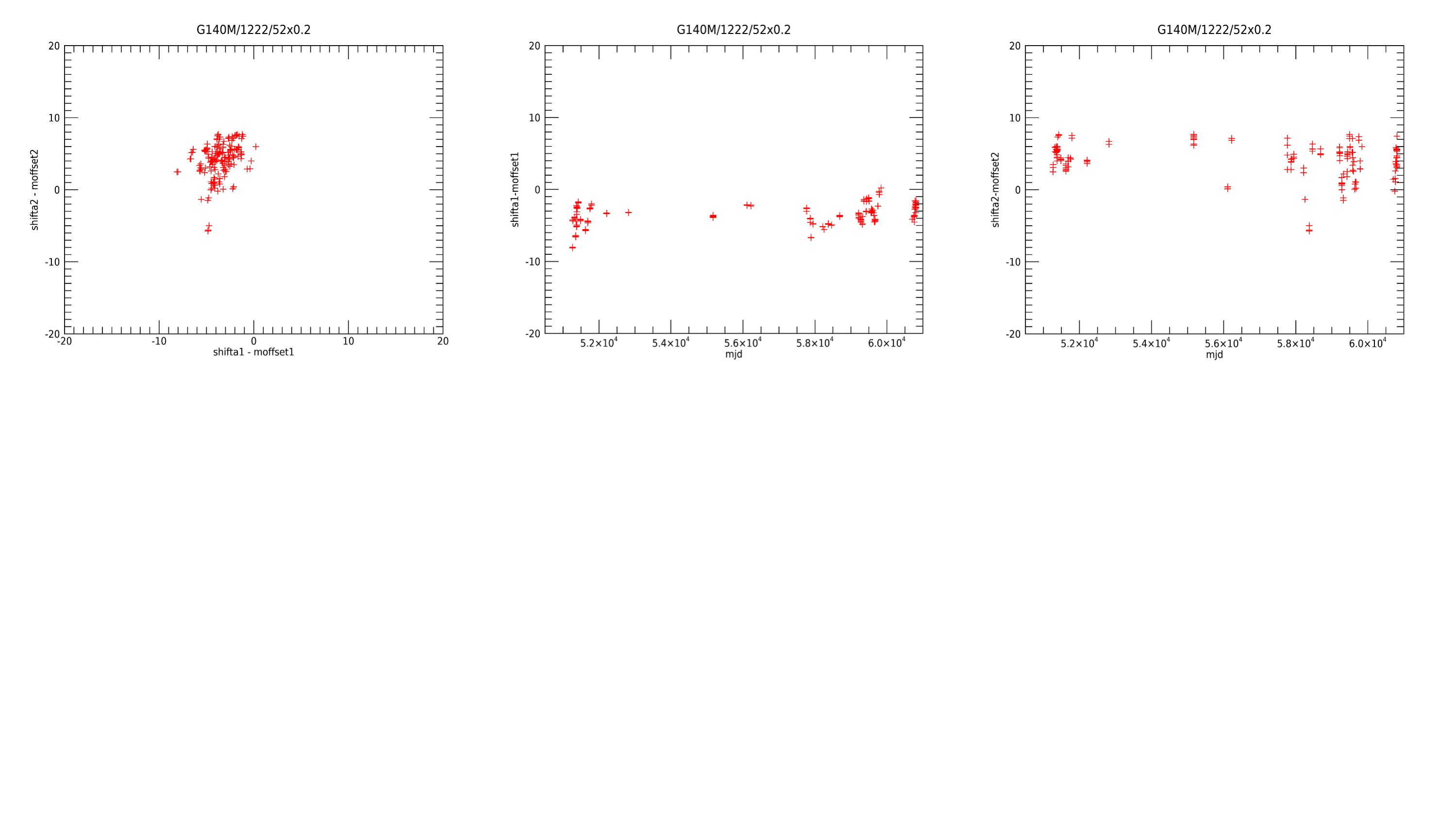}
    \caption{{\small SHIFTA $-$ MOFFSET} values (as in Fig.~\ref{fig:stats1}) for G140M/1222 wavecals taken through the 52x0.05, 52x0.1, and 52x0.2 apertures.
Note the expanded range of values (compared to the previous such plots) and the possible increase in scatter since MJD $\sim$58000 for 52x0.1.}
    \label{fig:stats9}
\end{figure}

Figures~\ref{fig:stats1}--\ref{fig:stats9} and Tables~\ref{tab:stats} and~\ref{tab:gstats} present the results of such an examination, for some of the more frequently used STIS FUV- and NUV-MAMA settings, based on STIS spectra available in the MAST archive through at least 2024July (MJD $\sim$ 60520).
In the figures, each set of three panels (left to right) shows {\small SHIFTA2$-$MOFFSET2} versus {\small SHIFTA1$-$MOFFSET1}, {\small SHIFTA1$-$MOFFSET1} versus MJD, and {\small SHIFTA2$-$MOFFSET2} versus MJD, for a particular combination of grating, central wavelength, and aperture.
We consider the difference between {\small SHIFTA} and the imposed MAMA offsets ({\small MOFFSET}), as the {\small SHIFTA} values are determined relative to template spectra which were not offset, and we wish to characterize the differences relative to the locations ``expected'' for the observations. In the central and right-hand panels, several temporal intervals can be defined, which can aid in understanding the distributions seen in the left-hand panels:
\begin{enumerate} 
\item{MJD $\sim$ 50700--53220 -- early STIS -- from the start of STIS operations to the suspension following the failure of the side 2 power supply. 
For much of this period (MJD $\sim$ 50825--52500), monthly offsets in y of up to $\pm$30 pix were performed for the echelle modes, via differences in the slit wheel and/or MSM values adopted for each grating/cenwave combination.}
\item{MJD $\sim$ 53220--55050 -- STIS inoperative, in safe mode.}
\item{MJD $\sim$ 55050--present -- post-SM4 (Servicing Mission 4) -- STIS operations restored.}
\end{enumerate}
Table~\ref{tab:stats} gives the overall means and standard deviations of the {\small SHIFTA$-$MOFFSET} values for the echelle modes shown in the figures, separated into the two periods (pre- and post-SM4) in which STIS has been in operation.
Table~\ref{tab:gstats} gives similar statistics for the G140L/1425 setting and for some of the shorter-wavelength G140M settings; Figure~\ref{fig:stats9} shows the {\small SHIFTA$-$MOFFSET} values obtained for G140M/1222 observations.
For the G140L and G140M modes, the monthly offsets were originally positive, but were changed to negative values in 1999Mar (MJD $\sim$ 51250) to avoid the FUV glow region of elevated background.
While the monthly offsets were discontinued for the echelle mode observations, they are still applied to observations with the various first-order MAMA modes.
The {\small SHIFTA2$-$MOFFSET2} values in Table~\ref{tab:gstats} and Figure~\ref{fig:stats9} are with reference to the ``base'' monthly offset, which is applied in December of each year.

Both general trends common to all the echelle and first-order settings and more specific features characterizing some of the individual settings can be discerned in the figures and tables:
\begin{itemize}
\item{In most of the left-hand panels, there appears to be a main cluster of points, often accompanied by one or more less-populated outlying sub-clusters and/or isolated ``discrepant'' points.
Examination of the MJD plots for the echelle settings suggests that the main cluster reflects a roughly constant (or at least slowly varying) behavior of the {\small SHIFTA$-$MOFFSET} values following the suspension of the monthly offsets (i.e., after MJD $\sim$ 52500, with similar behavior post-SM4).
The post-SM4 centroids of the main clusters can range from about $-$1 to +15 pix in x, and from about $-$10 to +10 pix in y, for the echelle settings listed in Table~\ref{tab:stats}.
Many of the sub-clusters appear to reflect the wider range of values seen while the monthly offsets were in effect -- perhaps related to the range of slit wheel and MSM values used to obtain the various offsets.}   
\item{For a given grating/cenwave combination, the mean {\small SHIFTA$-$MOFFSET} values appear to be fairly similar for different apertures (given the observed scatter and the relatively small number of datasets in some cases).}
\item{There are no obvious changes associated with the lamp changes for E140H/1234 and 1271 (at MJD $\sim$ 57785 and 58165, respectively) -- though relatively few observations have been taken at those settings since then.
(The high values of {\small SHIFTA2$-$MOFFSET2} for E140H/1234/0.2x0.2 for seven visits near MJD $\sim$ 58800 in Figure~\ref{fig:stats2} may not be representative, as they are for spectra of eta Car; see the next item.)}
\item{Some of the more obvious ``outliers'' in {\small SHIFTA2$-$MOFFSET2} may be associated with particular sight lines (or regions?).
For example, relatively recent observations of eta Car with E140H/1234 and E140H/1416 (though not with E140H/1598) appear to exhibit anomalously high {\small SHIFTA2$-$MOFFSET2} values.\footnotemark}
\footnotetext{Possibly related:  T. Gull (private communications in 2022-2025) has noted slight offsets in the wavelength zero points between E140H and E140M spectra of eta Car -- though the largest differences in {\scriptsize SHIFTA$-$MOFFSET} are for the direction perpendicular to the dispersion -- and has recalled early indications of echelle format shifting that may be related to changes in illumination of the spacecraft after some slews and/or after day/night crossings.}
\item{The {\small SHIFTA1$-$MOFFSET1} values for E140M/1425 exhibit a bi-modal distribution, seen most clearly post-SM4, with mean values separated by $\sim$2 pix (Fig.~\ref{fig:stats5}).
Those different values can be seen even for separate exposures within the same visit.}
\item{The one clear case where a wavecal yielded obviously spurious {\small SHIFTA} values (resulting in an erroneous wavelength solution) was for an E140H/1234 exposure (od9s02040) taken through the very small 0.1x0.03 aperture; see the post-SM4 entry for that setting in Table~\ref{tab:stats}.
In that case, it is hard to distinguish any lamp lines in the associated wavecal image -- even though it was taken not long after the switch to the HITM2 lamp.
The determination of the {\small SHIFTA} may have been misled by a hot pixel.}
\end{itemize}

For any given grating/cenwave combination, the scatter in the {\small SHIFTA$-$MOFFSET} values may reflect an inherent uncertainty in the actual positioning achieved for the corresponding set of slit wheel and MSM cylinder settings, together with possible thermally-induced shifts of the spectral format and any uncertainty in the cross-correlation of the individual wavecals with the template spectrum.
Given the limited number of observations for many of the settings, the overall scatter can also depend on the particular set of targets observed.
In an attempt to see if the scatter in {\small SHIFTA1$-$MOFFSET1} has increased over time (as the lamps have faded), Table~\ref{tab:stats1} lists the means and standard deviations for four time periods: pre-SM4 with monthly offsets, pre-SM4 without monthly offsets (for the echelle observations), post-SM4 before MJD = 57500 (roughly the halfway point of the post-SM4 period to date), and post-SM4 after MJD = 57500.
We will focus primarily on the last three of those periods (where the slit wheel and MSM values for each setting are most consistent), and on the shortest wavelength settings (where any effects of lamp fading would be expected to be most apparent). 
Examination of the figures and tables -- and of the detailed catalogs of exposure information on which the tables are based -- suggests that
\begin{itemize}
\item{For the echelle settings, the overall pre-SM4 standard deviations are generally larger than the post-SM4 values (due to the larger scatter during the period when monthly offsets were executed), but the pre-SM4 scatter seen after the offsets were discontinued is generally comparable to the post-SM4 values.}
\item{For most of the echelle settings, the overall post-SM4 scatter in {\small SHIFTA1$-$MOFFSET1} is well below 1 pix; the slightly larger value for E140M/1425 reflects the observed bi-modality.
The scatter in {\small SHIFTA2$-$MOFFSET2} is somewhat larger (more than 2 pix in some cases).
For a given echelle setting, the scatter among the {\small SHIFTA1$-$MOFFSET1} values within a single visit is typically even smaller -- less than several tenths of a pix -- suggesting that the accuracy of the wavelength zero point for any single observation generally should be well within the desired limit of 0.5-1.0 pix for MAMA observations.}
\item{The scatter for multiple observations of a single target with a given instrumental setting is often smaller than the overall value for that setting, particularly if the observations are obtained over a relatively short period of time.
This can significantly affect the overall standard deviations, for settings where most of the observations are of a very small number of targets.}
\item{The impression of increased scatter in {\small SHIFTA1$-$MOFFSET1} for G140M/1222/52x0.1 after MJD $\sim$58000 in Figure~\ref{fig:stats9} appears to be corroborated by the standard deviations for all three apertures listed in Table~\ref{tab:stats1}, which are $\sim$1.0 for the first half of the post-SM4 period and $\sim$1.5 for the second half.}
\item{While the standard deviations listed in Table~\ref{tab:stats1} for the E140H/1234 and E140H/1271 settings appear to be increasing slightly with time, the samples in each aperture/time bin are relatively small, the number of distinct targets in each bin is even smaller, and the differences in the standard deviations are likely within the mutual uncertainties.
There are no obvious trends in the standard deviations for the longer wavelength E140H or E230H settings.}
\end{itemize}
Given the limited number of observations for many of the shorter wavelength settings, the differences in target samples in the various aperture/time bins, and the number of factors that may contribute to the scatter, it is difficult to tell if the scatter characterizing those settings has increased significantly due to the fading of the wavecal lamps.

\begin{deluxetable}{ccccrrrrrr}
    \tabcolsep 4pt
    \tablewidth{0pt}
    \tablecaption{Statistics of {\small SHIFTA $-$ MOFFSET (E140 and E230)} \label{tab:stats}}
    \tabletypesize{\footnotesize}
    \tablehead{ \colhead{Setting} & \colhead{Aperture} & \colhead{Exptime} &
                \colhead{Number} & \colhead{Pre-SM4} & \colhead{Pre-SM4} &
                \colhead{N} & \colhead{Post-SM4} & \colhead{Post-SM4} &
                \colhead{N} \\
                   & & \colhead{(s)} & \colhead{ok/bad} & \colhead{Sha1$-$Mof1} & 
                \colhead{Sha2$-$Mof2} & & \colhead{Sha1$-$Mof1} & \colhead{Sha2$-$Mof2}}
    \startdata
E140H/1234 & 0.1x0.03 & 64 &  3/1 &   4.27(2.31) &   3.46(3.35) &  3 &
                                   [$-$171.47]   &[$-$226.50]   &  1 \\
           & 0.2x0.09 & 17 & 98/0 &   4.98(0.92) &$-$1.98(0.92) & 45 &
                                      5.54(0.56) &$-$1.70(2.26) & 53 \\
           & 0.2x0.2  & 13 & 20/0 &   3.52(1.64) &   0.06(3.37) &  8 &
                                      5.25(0.77) &   3.73(4.40) & 12 \\
E140H/1271 & 0.1x0.03 & 64 & 79/0 &   4.35(1.55) &$-$1.21(3.18) & 70 &
                                      5.56(0.36) &$-$1.65(1.10) &  9 \\
           & 0.2x0.09 & 17 & 91/0 &   3.64(1.46) &$-$0.38(2.39) & 43 &
                                      5.55(0.59) &$-$0.75(2.35) & 48 \\
           & 0.2x0.2  & 13 &157/0 &   3.72(2.11) &$-$2.06(1.68) & 66 &
                                      5.25(0.57) &$-$1.48(1.36) & 92 \\
E140H/1307 & 0.2x0.09 & 10 & 25/0 &   3.92(1.03) &   3.27(3.84) &  8 &
                                      8.49(0.50) &$-$2.44(0.98) & 17 \\
           & 0.2x0.2  & 10 & 75/0 &   5.51(0.00) &$-$3.87(0.00) &  1 &
                                      8.48(0.45) &$-$0.67(1.94) & 90 \\
E140H/1343 & 0.2x0.09 & 10 & 28/0 &   7.18(1.92) &   2.05(1.61) & 17 &
                                      7.87(0.84) &   2.28(1.09) & 11 \\
           & 0.2x0.2  & 10 & 57/0 &   5.80(1.65) &   8.78(3.44) & 14 &
                                      8.02(0.62) &   4.50(2.67) & 47 \\
E140H/1416 & 0.2x0.09 & 10 & 47/0 &   5.83(2.17) &$-$8.05(3.60) & 13 &
                                      7.91(0.46) &$-$9.85(1.93) & 34 \\
           & 0.2x0.2  & 10 & 99/0 &   6.36(1.36) &$-$9.50(1.06) & 18 &
                                      8.24(0.59) &$-$8.74(2.17) & 83 \\
E140H/1489 & 0.2x0.2  & 10 & 43/0 &   6.19(1.36) &   4.02(1.72) & 24 &
                                      8.73(0.61) &   5.93(0.90) & 19 \\
E140H/1598 & 0.2x0.09 & 10 & 25/0 &   0.57(1.86) &$-$0.52(2.99) & 16 &
                                      3.19(0.62) &$-$0.37(2.14) &  9 \\
           & 0.2x0.2  & 10 & 35/0 &   0.52(3.26) &$-$3.03(3.36) & 19 &
                                      2.73(0.81) &   0.47(1.22) & 16 \\
\hline
E140M/1425 & 0.2x0.06 & 12 &493/0 &   4.52(1.28) &   0.82(1.97) &273$^a$ &
                                                 &              &    \\
           &          &    &      &   3.75(0.55) &$-$0.92(0.68) &156$^b$ &
                                      3.73(1.01) &$-$0.46(0.82) & 66 \\
           & 0.2x0.2  & 10 &1622/0&   4.38(1.34) &   1.04(2.78) &621$^a$ &
                                                 &              &    \\
           &          &    &      &   3.95(0.66) &$-$0.78(0.79) &154$^b$ &
                                      4.11(0.81) &$-$0.33(0.80) &1015 \\
\hline
E230H/1763 & 0.1x0.03 & 36 & 51/0 &  13.92(1.27) &   3.08(2.26) & 48 &
                                     12.64(0.41) &$-$0.56(0.94) &  3 \\
           & 0.2x0.09 & 10 & 65/0 &  13.71(1.15) &$-$1.04(1.70) & 20 &
                                     13.24(0.40) &$-$1.53(0.98) & 45 \\
           & 0.2x0.2  & 10 & 20/0 &  12.32(1.40) &$-$0.30(2.91) & 16 &
                                     13.01(0.93) &$-$1.87(0.40) &  4 \\
E230H/1963 & 0.2x0.09 & 10 & 37/0 &   4.19(0.83) &   3.75(3.68) &  3 &
                                      5.88(0.39) &$-$4.71(2.22) & 34 \\
E230H/2013 & 0.2x0.09 & 10 &105/0 &  14.19(1.30) &   0.99(2.03) & 52 &
                                     14.53(0.37) &$-$1.63(0.59) & 53 \\
E230H/2063 & 0.2x0.09 & 10 & 33/0 &   5.12(0.35) &   3.17(1.00) &  4 &
                                      5.75(0.47) &$-$0.97(1.13) & 29 \\
E230H/2163 & 0.2x0.09 & 10 & 50/0 &   5.88(0.60) &   2.97(2.27) & 10 &
                                      5.65(0.45) &   0.20(1.01) & 40 \\
E230H/2263 & 0.2x0.09 & 10 & 60/0 &  13.21(1.85) &$-$5.76(2.87) & 10 &
                                     14.09(0.34) &$-$4.94(1.68) & 50 \\
           & 0.2x0.2  & 10 &107/0 &  13.07(1.10) &$-$3.20(2.95) & 30 &
                                     14.13(0.65) &$-$4.92(1.65) & 77 \\
E230H/2513 & 0.2x0.09 & 10 & 61/0 &$-$1.57(0.93) &$-$2.63(1.64) & 25 &
                                   $-$0.93(0.43) &$-$5.23(0.81) & 36 \\
           & 0.2x0.2  & 10 & 26/0 &$-$0.88(1.83) &$-$5.31(4.59) & 23 &
                                   $-$0.95(0.94) &$-$5.37(0.95) &  3 \\
E230H/2713 & 0.2x0.09 & 10 & 78/0 &   5.06(0.00) &  12.41(0.00) &  1 &
                                      5.30(0.40) &  10.39(1.09) & 77 \\
           & 0.2x0.2  & 10 &141/0 &   4.61(0.68) &  13.85(3.66) & 15 &
                                      5.32(0.52) &  10.17(0.71) &126 \\
\hline
E230M/1978 & 0.2x0.06 & 10 & 78/0 &   0.47(0.73) &   0.42(1.85) & 23$^a$ &
                                                 &              &    \\
           &          &    &      &$-$1.11(0.46) &$-$2.40(0.50) & 11$^b$ &
                                   $-$0.81(0.33) &$-$2.63(0.46) & 44 \\
           & 0.2x0.2  & 10 &378/0 &   0.36(0.83) &$-$0.20(2.26) & 85$^a$ &
                                                 &              &    \\
           &          &    &      &$-$0.84(0.41) &$-$1.96(0.89) & 23$^b$ &
                                   $-$0.51(0.33) &$-$2.50(0.63) &270 \\
E230M/2707 & 0.2x0.06 & 10 &240/0 &   5.22(2.31) &   1.33(1.42) & 77$^a$ &
                                                 &              &    \\
           &          &    &      &   3.34(0.41) &$-$1.02(0.82) & 62$^b$ &
                                      3.57(0.40) &$-$1.45(1.08) &101 \\
           & 0.2x0.2  & 10 &368/0 &   4.07(1.59) &   0.86(1.54) &106$^a$ &
                                                 &              &    \\
           &          &    &      &   3.41(0.36) &$-$1.81(0.95) & 40$^b$ &
                                      3.65(0.43) &$-$1.14(1.03) &222 \\
    \enddata
\tablecomments{Values for {\sc shifta}$-${\sc moffset} are mean(standard deviation). 
$^a$ = monthly offsets; $^b$ = no monthly offsets; all other pre-SM4 combine both.
The {\sc shifta} values are taken from the science exposure headers, and are average values for the (relatively few) cases where wavecals were taken both before and after.}
\end{deluxetable}

\begin{deluxetable}{lcccrrr}
    \tabcolsep 4pt
    \tablewidth{0pt}
    \tablecaption{Statistics of {\small SHIFTA $-$ MOFFSET (G140)} \label{tab:gstats}}
    \tabletypesize{\footnotesize}
    \tablehead{ \colhead{Setting} & \colhead{Aperture} & \colhead{Mof2} &
                \colhead{Period} & \colhead{Sha1$-$Mof1} & \colhead{Sha2$-$Mof2} &
                \colhead{N}} 
    \startdata
G140L/1425 & 52x0.1    & $-$134 & all      &$-$2.84(1.12) &$-$5.39(3.93) &  64 \\
           & 52x0.2    &        & all      &$-$2.91(1.40) &$-$4.90(5.59) &1111 \\
           & 52x0.5$^a$&        & all      &$-$2.63(1.28) &$-$4.77(4.67) & 631 \\
           & 52x2$^a$  &        & all      &$-$2.10(1.46) &$-$4.37(4.68) & 616 \\
\hline
G140M/1173 & 52x2$^b$  & $-$109 & pre-SM4  &$-$0.10(0.70) &  10.20(1.57) &  25 \\
           &           &        & post-SM4 &   0.28(0.68) &   9.74(1.49) &  14 \\
G140M/1218 & 52x0.05   & $-$121 & post-SM4 &   0.08(1.56) &   4.70(3.32) &  15 \\
           & 52x0.2    &        & all      &$-$0.21(1.72) &   4.80(1.88) &  16 \\
G140M/1222 & 52x0.05   & $-$109 & post-SM4 &$-$3.26(1.44) &   5.43(4.98) & 106 \\
           & 52x0.1    &        & all      &$-$3.36(1.37) &   4.86(3.64) & 243 \\
           & 52x0.2    &        & all      &$-$3.59(1.49) &   4.68(3.91) & 156 \\
           & 52x0.5$^b$&        & all      &$-$3.57(0.76) &   5.21(3.04) &  22 \\
           & 52x2$^b$  &        & all      &$-$3.30(2.41) &   3.90(3.30) &  12 \\
G140M/1272 & 52x0.1    & $-$117 & pre-SM4  &$-$6.94(1.34) &$-$5.50(2.67) &  18 \\
           & 52x0.2    &        & pre-SM4  &$-$4.59(2.81) &$-$1.25(3.85) &  24 \\
           & 52x2$^b$  &        & all      &$-$5.94(1.01) &$-$4.69(2.06) &  26 \\
G140M/1321 & 52x0.1    & $-$117 & pre-SM4  &   0.42(1.29) &$-$5.03(2.20) &  17 \\
           & 52x0.2    &        & all      &   0.59(1.66) &$-$5.00(3.04) &  13 \\
           & 52x0.5$^b$&        & pre-SM4  &   0.40(0.92) &$-$4.40(1.74) &  31 \\
    \enddata
\tablecomments{Values for {\sc shifta}$-${\sc moffset} are mean(standard deviation);
Mof2 values are the ``base'' {\sc moffset2} values (used in December). 
$^a$ = wavecals used 52x0.05 aperture; $^b$ = wavecals used 52x0.1 aperture.
The {\sc shifta} values are taken from the science exposure headers, and are average values for the (relatively few) cases where wavecals were taken both before and after.}
\end{deluxetable}

\begin{deluxetable}{ccrrrrrrrr}
    \tabcolsep 4pt
    \tablewidth{0pt}
    \tablecaption{Statistics of {\small SHIFTA1 $-$ MOFFSET1} \label{tab:stats1}}
    \tabletypesize{\footnotesize}
    \tablehead{ \colhead{Setting} & \colhead{Aperture} & \colhead{Pre Mof$^a$} &
                \colhead{N} & \colhead{Pre No Mof$^b$} & \colhead{N} & \colhead{Post A$^c$} &
                \colhead{N} & \colhead{Post B$^d$} & \colhead{N}}
    \startdata
E140H/1234 & 0.2x0.09 &  4.59(1.23) & 19 &  5.26(0.45) & 26 &
                         5.58(0.53) & 34 &  5.48(0.61) & 19 \\
E140H/1271 & 0.1x0.03 &  4.03(1.62) & 54 &  5.43(0.55) & 16 &
                         \nodata    &  0 &  \nodata    &  0 \\
           & 0.2x0.09 &  3.09(1.26) & 32 &  5.25(0.48) & 11 &
                         5.60(0.54) & 37 &  5.41(0.73) & 11 \\
           & 0.2x0.2  &  3.56(2.19) & 58 &  4.93(0.43) &  8 &
                         5.39(0.50) & 61 &  4.97(0.61) & 31 \\
E140H/1307 & 0.2x0.2  &  5.51(0.00) &  1 &  \nodata    &  0 &
                         8.40(0.69) & 11 &  8.50(0.41) & 79 \\
E140H/1343 & 0.2x0.09 &  6.51(2.95) &  7 &  7.65(0.40) & 10 &
                         7.89(0.88) & 10 &  7.66(0.00) &  1 \\
           & 0.2x0.2  &  5.69(1.66) & 13 &  7.19(0.00) &  1 &
                         8.43(0.97) &  7 &  7.95(0.52) & 40 \\
E140H/1416 & 0.2x0.09 &  5.46(2.16) & 11 &  7.85(0.49) &  2 &
                         7.77(0.52) & 20 &  8.11(0.39) & 14 \\
           & 0.2x0.2  &  5.67(1.08) & 12 &  7.73(0.58) &  6 &
                         8.24(0.52) & 30 &  8.24(0.63) & 53 \\
\hline
E140M/1425 & 0.2x0.06 &  4.52(1.28) &273 &  3.75(0.55) &156 &
                         3.48(1.16) & 34 &  4.01(0.75) & 32 \\
           & 0.2x0.2  &  4.38(1.34) &621 &  3.95(0.66) &154 &
                         3.93(0.98) &375 &  4.21(0.67) &640 \\
\hline
E230H/1763 & 0.2x0.09 & 14.28(1.16) & 12 & 12.87(0.31) &  8 &
                        13.51(0.44) &  6 & 13.20(0.39) & 39 \\
E230H/2013 & 0.2x0.09 & 14.19(1.61) & 32 & 14.13(0.54) & 20 &
                        14.56(0.00) &  1 & 14.53(0.37) & 52 \\
E230H/2263 & 0.2x0.09 & 12.27(2.86) &  4 & 13.84(0.21) &  6 &
                        14.04(0.37) & 31 & 14.17(0.30) & 19 \\
           & 0.2x0.2  & 12.88(1.37) & 18 & 13.36(0.41) & 12 &
                        13.82(0.52) & 29 & 14.32(0.65) & 48 \\
E230H/2513 & 0.2x0.09 &$-$1.62(1.00)& 21 &$-$1.33(0.42)&  4 &
                       $-$0.91(0.50)& 25 &$-$0.98(0.23)& 11 \\
E230H/2713 & 0.2x0.09 &  5.06(0.00) &  1 &  \nodata    &  0 &
                         5.28(0.33) & 36 &  5.32(0.45) & 41 \\
           & 0.2x0.2  &  4.67(0.85) &  9 &  4.51(0.37) &  6 &
                         5.21(0.47) & 78 &  5.49(0.55) & 48 \\
\hline
E230M/1978 & 0.2x0.2  &  0.36(0.83) & 85 &$-$0.84(0.41)& 23 &
                       $-$0.69(0.28)&113 &$-$0.38(0.31)&157 \\
E230M/2707 & 0.2x0.06 &  5.22(2.31) & 77 &  3.34(0.41) & 62 &
                         3.60(0.43) & 85 &  3.41(0.12) & 16 \\
           & 0.2x0.2  &  4.07(1.59) &106 &  3.41(0.36) & 40 &
                         3.66(0.33) & 85 &  3.65(0.49) &137 \\
\hline
G140M/1222 & 52x0.05  &  \nodata    &  0 &  \nodata    &  0 &
                       $-$3.78(1.07)& 52 &$-$2.76(1.57)& 54 \\
           & 52x0.1   &$-$3.60(1.08)& 52 &  \nodata    &  0 &
                       $-$3.24(0.83)& 47 &$-$3.54(1.68)& 93 \\
           & 52x0.2   &$-$4.19(1.52)& 67 &  \nodata    &  0 &
                       $-$3.30(0.73)& 14 &$-$3.29(1.42)& 93 \\
    \enddata
\tablecomments{Values for {\sc shifta1}$-${\sc moffset1} are mean(standard deviation).
$^a$ = pre-SM4, with monthly offsets; $^b$ = pre-SM4, with no monthly offsets; $^c$ = post-SM4 (MJD $<$ 57500); $^d$ = post-SM4 (MJD $>$ 57500).
The {\sc shifta} values are taken from the science exposure headers, and are average values for the (relatively few) cases where wavecals were taken both before and after.}
\end{deluxetable}

\clearpage
\ssection{Simulations}\label{sec:sims}

Prior to the switch of the GO wavecal lamp for the E140H/1234 and 1271 settings, long TIME-TAG exposures were obtained with the HITM2 lamp at those settings (Cycle 23 PID 14489; executed 2016Mar), in order to gauge the exposure times that would be needed (\href{https://www.stsci.edu/files/live/sites/www/files/home/hst/instrumentation/stis/documentation/instrument-science-reports/_documents/2017_04.pdf}{Peeples 2017}).
Random samplings of the exposure for E140H/1234 (1000s through the 0.2x0.09 aperture) suggested that an exposure time of 16s would be sufficient to achieve the desired accuracy in the wavelength zero point, and so a default exposure time of 17s was adopted for that setting/aperture combination.

\begin{figure}[!h]
  \centering
  \includegraphics[width=150mm]{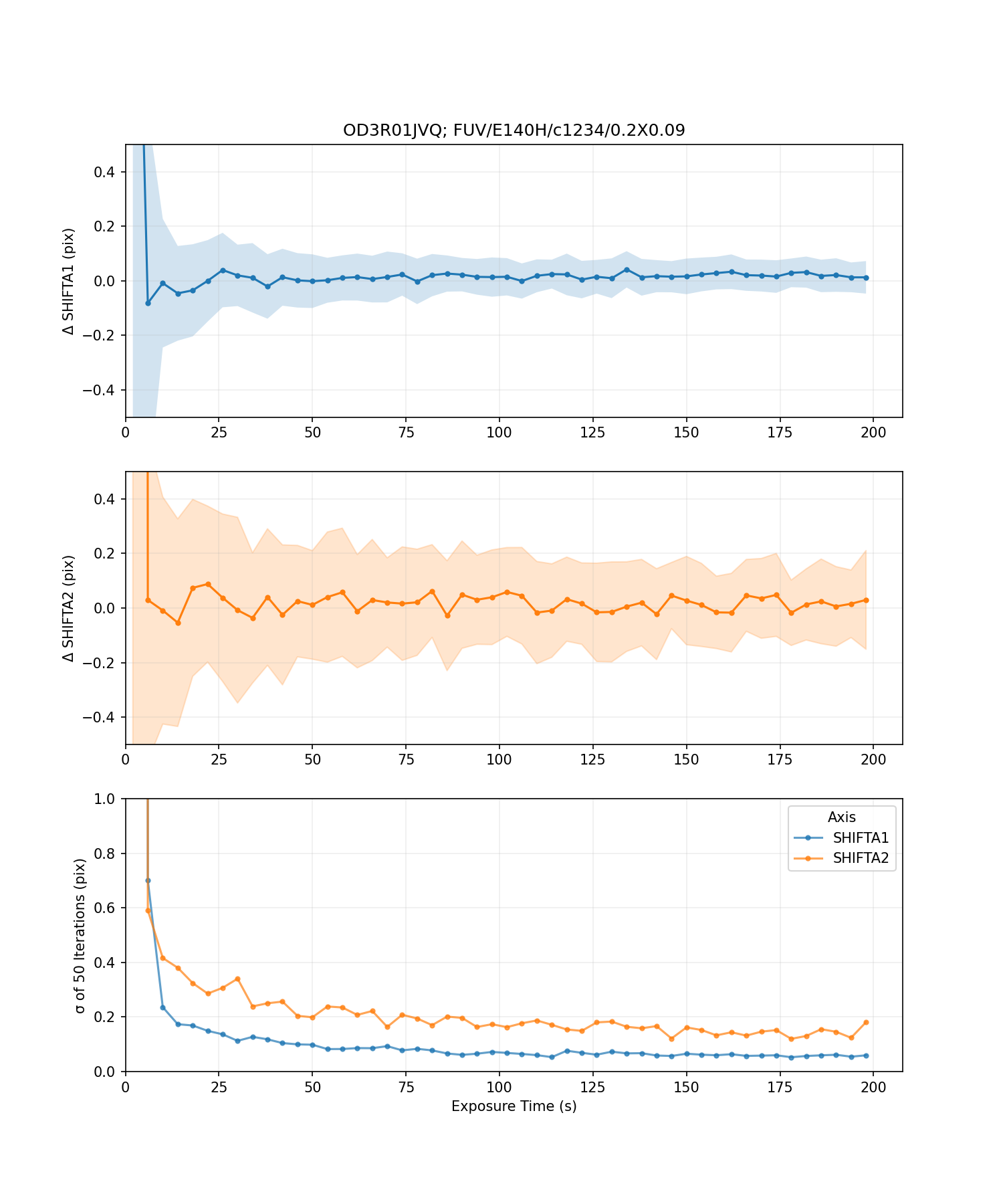}
    \caption{{\small SHIFTA1} (top) and {\small SHIFTA2} (middle) values obtained from sub-sampling a time-tagged 1000-sec E140H/1234 HITM2 wavecal obtained in 2016Mar through the 0.2x0.09 aperture (PID 14489).
For each exposure time, 50 independent samplings are performed; the solid lines show the average {\small SHIFTA} values, while the shaded region shows the 1$\sigma$ deviations (which are also shown in the bottom panel). 
Fairly consistent {\small SHIFTA} values, with scatter less than about 0.15 pix in x and 0.3 pix in y, are obtained for sub-exposure times greater than about 25 sec (vs. the default 17 sec) -- but the lamp has continued to fade since then.}
    \label{fig:sims}
\end{figure}

\begin{figure}[!h]
  \centering
  \includegraphics[width=170mm]{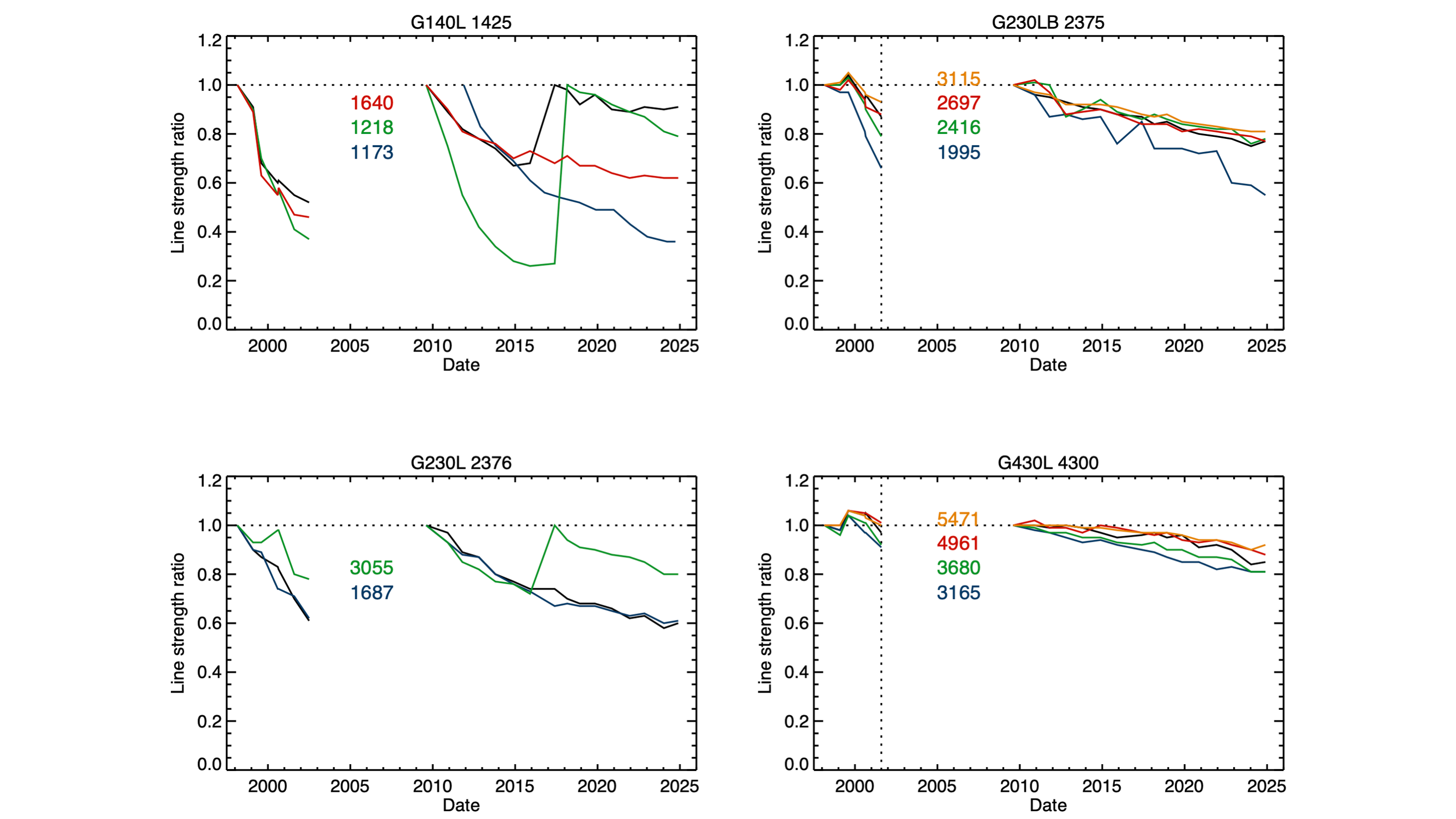}
    \caption{Weighted average relative strengths of the emission lines in the Pt/Cr-Ne lamps used for wavelength calibration, for selected first-order G140L/M (FUV-MAMA), G230L/M (NUV-MAMA), G230LB/MB (CCD), and G430L/M (CCD) settings.  Ratios before 2003 are relative to the initial cycle 7 line strengths.  Ratios after 2009 (post-SM4) are relative to cycle 17.  Other jumps in the ratios indicate where the lamp and/or the lamp current was changed to obtain more measurable lines.  All three of the Pt/Cr-Ne lamps are fading, most severely at the shortest wavelengths.}
    \label{fig:ratios3}
\end{figure}

We have conducted similar simulations, based on that same long E140H/1234 exposure.
For total exposure times ranging from 6s to 198s, {\small SHIFTA} values were obtained in the usual way (2D cross-correlation against a template spectral image) for 50 random sub-samples of the time-tagged wavecal data.
In Figure~\ref{fig:sims}, the top two panels show the mean$\pm$1$\sigma$ deviation in {\small SHIFTA1} and {\small SHIFTA2}, as functions of exposure time, for those 50-element sets; the bottom panel shows the 1$\sigma$ deviations themselves.
The simulations indicate that mean {\small SHIFTA} values very close to those derived from the full 1000s exposure -- with 1$\sigma$ deviations of order 0.15 pix in {\small SHIFTA1} and 0.3 pix in {\small SHIFTA2} -- could have been obtained (in 2016) for exposure times greater than about 25s.
We also added random sampling from a long FUV dark exposure (oe6ae6toq) at the simulated exposure times, taken when the detector had been on for over 6 hours, in order to gauge the effects of significantly higher dark counts.  
For exposure times greater than 14s, similar results to the low-dark case were obtained.
Since 2016, wavecals for the G140M/1173 setting, obtained annually in one of the regular STIS calibration programs (and covering the short wavelength end of the E140H/1234 setting), suggest that the HITM2 lamp has faded by another factor of $\sim$1.6 at those short wavelengths (Fig.~\ref{fig:ratios3})-- so the corresponding current (2025) minimum exposure time for E140H/1234/0.2x0.09 would be $\sim$40s.

\clearpage
\ssection{Conclusions and Recommendations}\label{sec:concl}

\begin{figure}[!h]
  \centering
  \includegraphics[width=170mm]{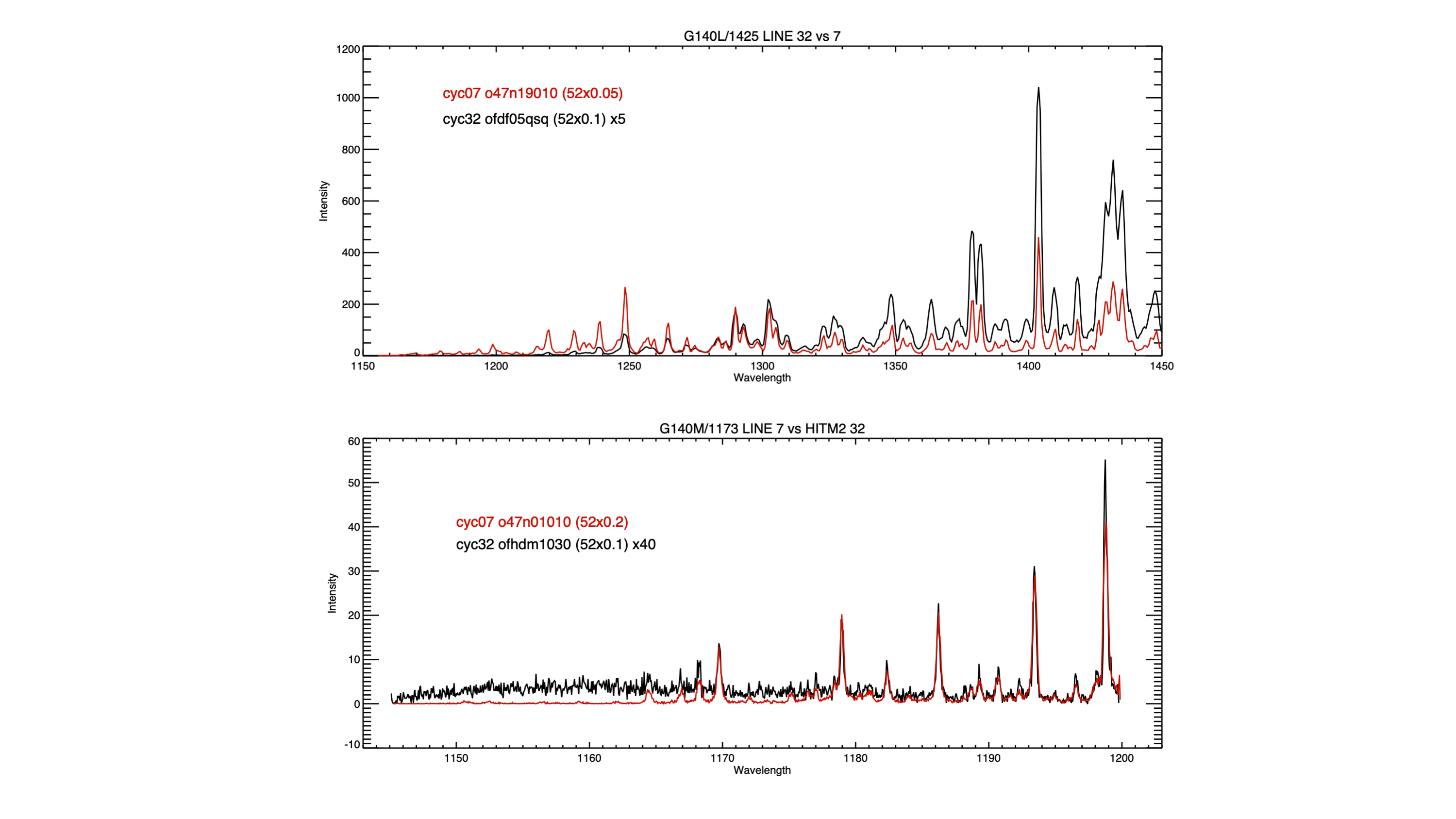}
    \caption{({\it upper}) G140L/1425 LINE lamp spectra from cycle 7 (taken through the 52x0.05 aperture; red) and cycle 32 (taken through the 52x0.1 aperture; black).  
The cycle 32 spectrum has been multiplied by 5 (or by 10, in effect, given the difference in apertures) to compensate somewhat for the fading of the lines. 
The LINE lamp thus has faded by more than a factor of 100 below 1200 \AA, by a factor of $\sim$30 near 1250 \AA, by a factor of $\sim$10 near 1290 \AA, and by a factor of 4-5 above 1400 \AA, between 1997 and 2024.
({\it lower}) G140M/1173 lamp spectra from cycle 7 (LINE lamp; 52x0.2 aperture; red) and from cycle 32 (HITM2 lamp; 52x0.1 aperture; black).  
The cycle 32 spectrum has been multiplied by 40 (or by 20, in effect, given the difference in apertures) to compensate for the fading of the lines. 
In this spectral range, the HITM2 lines are currently a factor of $\sim$20 weaker than those from the LINE lamp in cycle 7.}
    \label{fig:g140}
\end{figure}

Because the lamps have faded most dramatically at wavelengths below about 1300 \AA, the settings most vulnerable to inaccurate (or failed) {\small SHIFTA} determinations are the ones at those shorter wavelengths -- especially those with limited spectral coverage, those at higher spectral resolutions, and/or those using the smallest apertures.
We therefore focus on the shortest wavelength E140H and G140M settings. 
(The E140M/1425 and G140L/1425 settings cover much larger wavelength ranges, which include many reasonably strong lines that are less affected by the fading.)
As noted above, the initial wavecal exposure times appear to have been fairly generous, given that reasonable {\small SHIFTA} values continue to be obtained for line strengths of order 5-10\% of the original values.
While it therefore does not seem necessary to obtain the total counts seen in the lines in the initial wavecals, we would recommend a target of $\sim$20\% of the original total LINE lamp counts for the echelle settings (so, e.g., increasing the exposure time by a factor of 4 if the lines are now weaker by a factor of 20).
Such an increase should enable reliable {\small SHIFTA} values to be determined even as the lamps continue to fade.
\begin{itemize}
\item{E140H/1234/HITM2 -- The simulations discussed above suggest a minimum exposure time of 40s (roughly 2.5x the current default) when using the 0.2x0.09 aperture.  
The one failed wavecal was for this setting -- in 2017, with the 0.1x0.03 aperture, and with an exposure time corresponding to the current default for that aperture.  
A cycle 32 HITM2 wavecal for G140M/1173 (which covers the short wavelength end of the E140H/1234 setting) indicates that the HITM2 lines near 1200 \AA\ are weaker by a factor $\sim$20 than those seen initially for the LINE lamp (for that same setting); see Figure~\ref{fig:g140}.  
We therefore recommend increasing the current default exposure times for this setting by a factor $\sim$4.}
\item{E140H/1271/HITM2 -- The lamp lines have faded less at the slightly longer wavelengths covered by this frequently used setting (weaker by less than a factor $\sim$15 than the original LINE lamp), so an increase by a factor $\sim$2.5 should suffice.}
\item{E140H/1307/LINE -- The LINE lamp still has stronger lines than the HITM lamps at these slightly longer wavelengths; an increase by a factor $\sim$1.5 should suffice (see Appendix A).}
\end{itemize}
For most of the first-order settings (except 1173), we compare the current line strengths with the initial HITM1 line strengths.  
Because the HITM1 lamp has faded much less than the LINE lamp -- by factors of about 8 and 3 near 1200 and 1250 \AA, respectively -- these shorter wavelength settings should require at most modest increases (factors of 1.5-2).

\begin{itemize}
\item{G140M/1173/HITM2 -- This setting is used in the annual MAMA spectroscopic sensitivity monitor (and the wavecal data since 2012 have been incorporated into the annual dispersion monitor), but it has not been used in GO programs for many years.
The wavecal lamp was changed from LINE to HITM2 in 2012.
While the lines have weakened by a factor $\sim$2.5 since then (Fig.~\ref{fig:ratios3}), and fewer lines are now detectable, there are still a few well-determined lines visible in the 2D images -- which should be sufficient for {\small SHIFTA} determinations.}
\item{G140M/1218,1222/HITM1 -- The 1218 setting is little used; the last wavecals obtained (in 2020) show a number of measurable lines, so no changes appear to be needed.
The nearly coincident 1222 setting, however, continues to be fairly heavily used, e.g., for obtaining spectra of Lyman-$\alpha$ emission from late-type stars.
This setting includes a fairly strong line near 1248 \AA\ that is not covered by the 1218 setting, and recent default wavecals appear to yield reasonable {\small SHIFTA} values, but it might be prudent, in view of the possible increased scatter in {\small SHIFTA1$-$MOFFSET1}, to increase the default exposure time to be similar to that for 1218.}
\item{G140M/1272/HITM1 -- As the lamp has faded less at these slightly longer wavelengths, increasing the exposure time by a factor $\sim$1.5 should be sufficient.}
\end{itemize}
Table~\ref{tab:texp} summarizes the current default exposure times and our recommendations for increased exposure times for these shortest wavelength E140H and G140M settings, for the apertures used for the wavecals.
Until the default exposure times can be changed, these recommendations may be implemented in GO wavecals, in consultation with the STIS contact scientists.
Switching the lamp for the G140M settings from the current default HITM1 to the brighter HITM2 could reduce the exposure times by a factor $\sim$2.
Given the typical post-SM4 usage of those settings, the proposed increases in exposure time for the default wavecals would add $\sim$1.7 mA hr per year for the E140H settings and $\sim$6.6 mA hr per year for the G140M settings -- corresponding to small fractional increases in the current typical lamp usages (Table~\ref{tab:usage}).
 
\begin{deluxetable}{cccccc}
    \tabcolsep 4pt
    \tablewidth{0pt}
    \tablecaption{Current / Suggested Exposure Times \label{tab:texp}}
    \tabletypesize{\footnotesize}
    \tablehead{ \colhead{Setting} & \colhead{Lamp} & \colhead{Coverage} & 
                \colhead{0.1x0.03} & \colhead{0.2x0.09} & \colhead{0.2x0.2} }
    \startdata
E140H/1234 & HITM2 & 1142-1335 & 64s / 250s & 17s / 70s & 13s / 50s \\
E140H/1271 & HITM2 & 1161-1361 & 64s / 150s & 17s / 40s & 13s / 30s \\
E140H/1307 & LINE  & 1197-1397 & 33s / 50s  & 10s / 15s & 10s / 15s \\
\hline
Setting    & Lamp  & Coverage  & 52x0.05    & 52x0.1    & 52x0.2    \\
\hline
G140M/1173 & HITM2 & 1145-1201 & 188s / nc  & 150s / nc  & 150s / nc \\
G140M/1218 & HITM1 & 1190-1245 & 139s / nc  & 111s / nc  & 111s / nc \\
G140M/1222 & HITM1 & 1194-1249 & 72s / 140s & 58s / 110s & 58s / 110s\\
G140M/1272 & HITM1 & 1245-1299 & 82s / 120s & 66s / 100s & 66s / 100s\\
    \enddata
\end{deluxetable}



\clearpage
\vspace{-0.3cm}
\ssectionstar{Acknowledgements}
\vspace{-0.3cm}
We thank A. Welty for providing information on the default wavecal lamps and exposure times, T. Gull for comments regarding early investigations of the wavelength zero point accuracy, and M. Mingozzi for a careful reading of the earlier versions of this report.

\vspace{-0.3cm}
\ssectionstar{Change History for STIS ISR 2025-05}\label{sec:History}
\vspace{-0.3cm}
Version 1: \ddmonthyyyy{18 September 2025} - Original Document 

\vspace{-0.3cm}
\ssectionstar{References}\label{sec:References}
\vspace{-0.3cm}

\noindent
\href{https://www.stsci.edu/files/live/sites/www/files/home/hst/instrumentation/stis/documentation/instrument-science-reports/_documents/199701.pdf}{Baum, S. A. 1997}, STIS Instrument Science Report 1997-01, Automatic and GO Wavecals, for CCD and MAMA Spectroscopic Observations

\noindent
\href{https://www.stsci.edu/files/live/sites/www/files/home/hst/instrumentation/stis/documentation/instrument-science-reports/_documents/199812.pdf}{Hodge, P., Baum, S, McGrath, M., et al. 1998}, STIS Instrument Science Report 1998-12, Calstis4, calstis11, calstis12: Wavecal Processing in the STIS Calibration Pipeline

\noindent
\href{https://www.stsci.edu/files/live/sites/www/files/home/hst/instrumentation/stis/documentation/instrument-science-reports/_documents/199619.pdf}{Hulbert, S, Hodge, P., \& Baum, S. 1996}, STIS Instrument Science Report 1996-19, The STScI STIS Pipeline VI: Reduction of WAVECALs

\noindent
\href{https://ui.adsabs.harvard.edu/abs/2004SPIE.5488..679K/abstract}{Kerber, F., Rosa, M. R., Sansonetti, C. J., et al., 2004}, Proc. SPIE, vol. 5488, 679, Spectral Characterization of HST Calibration Lamps -- New Pt/Cr-Ne line catalogues and ageing test

\noindent
\href{https://ui.adsabs.harvard.edu/abs/2006hstc.conf..318K/abstract}{Kerber, F., Bristow, P., Rosa, M. R., et al. 2006}, NASA CP2006-214134, 318, Characterization of Pt/Cr-Ne Hollow-Cathode Lamps for Wavelength Standards in Space Astronomy

\noindent
Lindler, D., Kimble, R., Plait, P., et al. 1996, STIS pre-launch analysis report r0013, STIS Automatic WAVECALs

\noindent
\href{https://www.stsci.edu/files/live/sites/www/files/home/hst/instrumentation/stis/documentation/instrument-science-reports/_documents/2022_07.pdf}{Pascucci, I., Hodge, P., Proffitt, C. R., \& Ayres, T. 2011}, STIS Instrument Science Report 2011-01, Wavelength Calibration Accuracy for the STIS CCD and MAMA Modes (Pas11)

\noindent
\href{https://www.stsci.edu/files/live/sites/www/files/home/hst/instrumentation/stis/documentation/instrument-science-reports/_documents/2022_07.pdf}{Peeples, M. 2017}, STIS Instrument Science Report 2017-04, On the Fading of the STIS Ultraviolet Calibration Lamps 

\noindent
\href{https://www.stsci.edu/files/live/sites/www/files/home/hst/instrumentation/stis/documentation/instrument-science-reports/_documents/2022_07.pdf}{Sonnentrucker, P. 2015}, STIS Instrument Science Report 2015-02, Multi-Cycle Analysis of the STIS Dispersion Solutions

\noindent
\href{https://www.stsci.edu/files/live/sites/www/files/home/hst/instrumentation/stis/documentation/instrument-science-reports/_documents/2018_04.pdf}{Welty, D. E. 2018}, STIS ISR 2018-04, Monitoring the STIS Wavelength Calibration: MAMA and CCD First-Order Modes

\clearpage
\ssectionstar{Appendix A -- Wavelength calibration lamps}\label{sec:AppA}
STIS has three internal Pt/Cr-Ne hollow cathode lamps (LINE, HITM1, HITM2) that are used for the dispersion monitors, the slit wheel repeatability monitor, the routine wavecals obtained with most STIS spectroscopic exposures, and aperture location during target acquisitions.
All three lamps have faded over time, exhibiting a roughly constant fractional decline per year (Fig.~\ref{fig:ratios3}; see \href{https://www.stsci.edu/files/live/sites/www/files/home/hst/instrumentation/stis/documentation/instrument-science-reports/_documents/2017_04.pdf}{Peeples 2017}, \href{https://www.stsci.edu/files/live/sites/www/files/home/hst/instrumentation/stis/documentation/instrument-science-reports/_documents/2018_04.pdf}{Welty 2018}, and the plots available on the \href{https://www.stsci.edu/hst/instrumentation/stis/performance/monitoring}{monitors web page}).
The fading has been most severe at the shortest wavelengths, and more severe for the LINE lamp than for the HITM lamps.
Figure~\ref{fig:g140} compares LINE lamp spectra obtained in cycle 7 (o47n19010, 10 mA, 52x0.05; in red) and in cycle 32 (ofdf05qsq, 10 mA, 52x0.1; multiplied by 5; in black).
Multiplication of the cycle 32 spectrum by 5 has the effect of comparing 10 times the current LINE spectrum with the cycle 7 LINE spectrum, for a common aperture.
The LINE lamp has faded by more than a factor of 100 below 1200 \AA, by a factor of $\sim$30 near 1250 \AA, by a factor of $\sim$10 near 1290 \AA, and by a factor of 4-5 above 1400 \AA.
As noted in the main body of this report, these changes in lamp brightness have necessitated changes in the exposure time, lamp current, and/or specific lamp used for some of the calibration observations and wavecals.

\ssectionstar{Appendix B -- Pt/Cr-Ne lamp usage}\label{sec:AppB}
A detailed record of the use of the various STIS calibration lamps, from STIS installation in 1997 through the end of 2018, had been maintained by T. Wheeler.
Table~\ref{tab:usage} gives a summary of the total usage (in mA hr) of the HITM1 and LINE lamps -- including pre-launch ground testing and the yearly in-orbit totals for 1997 through 2018.
To estimate the total usage as of early 2025, we assume a usage during 2019--2024 of 90 mA hr per year for each of those two lamps (roughly the post-SM4 maxima). 
The total usage would then be about 2830 mA hr for the HITM1 lamp and about 3480 mA hr for the LINE lamp -- or about 19\% and 23\% of the ``expected'' 15000 mA hr lifetimes for the two lamps (e.g., \href{https://ui.adsabs.harvard.edu/abs/2004SPIE.5488..679K/abstract}{Kerber et al. 2004}).
The HITM2 lamp has been used much less -- only about 160 mA hr in ground testing and 110 mA hr in in-orbit usage (as of the end of 2018) -- though it is now being used for calibrations and wavecals for several of the shorter-wavelength settings.

\begin{deluxetable}{lrr}
    \tabcolsep 4pt
    \tablewidth{0pt}
    \tablecaption{Pt/Cr-Ne Lamp Usage \label{tab:usage}}
    \tabletypesize{\footnotesize}
    \tablehead{ \colhead{Year} & \colhead{HITM1} & \colhead{LINE} \\
                               & \colhead{(mA hr)} & \colhead{(mA hr)} }
    \startdata
Pre-launch &  538.0 &  796.0 \\
\hline
1997$^a$   &   68.5 &  182.7 \\
1998       &  104.8 &  136.7 \\
1999       &  178.9 &  172.2 \\
2000       &  167.6 &  257.6 \\
2001       &  188.9 &  177.0 \\
2002       &  128.7 &  222.6 \\
2003       &  168.5 &  135.7 \\
2004$^a$   &   56.4 &   86.7 \\
\hline
2009$^a$   &   40.5 &   82.0 \\
2010       &   71.0 &  143.4 \\
2011       &   72.6 &   87.0 \\
2012       &   80.6 &   47.2 \\
2013       &   57.6 &   69.1 \\
2014       &   52.0 &   77.3 \\
2015       &   64.4 &   76.1 \\
2016       &   72.2 &   56.1 \\
2017       &   87.8 &   67.0 \\
2018       &   88.9 &   65.1 \\
\hline
total      & 2288.3 & 2938.4 \\
    \enddata
\tablecomments{$^a$ = partial year}
\end{deluxetable}

\end{document}